\documentclass{aa}
\usepackage{natbib}

\usepackage{graphicx}
\usepackage{txfonts}
\usepackage{lipsum}
\usepackage{subcaption}         
\usepackage{lscape}
\usepackage{placeins}
\usepackage{amsmath}

\newcommand{\matr}[1]{\mathbf{#1}}
\usepackage[colorlinks=true,linkcolor=blue,citecolor=blue,urlcolor=blue]{hyperref}

\begin{document}

   \title{The correlation between voids identified in 3D large-scale-structure and 2D weak-lensing maps}

   \author{Y. Fang\inst{1, 2}\thanks{E-mail: \url{yuedong.fang@duke.edu}}
        \and C. T. Davies\inst{1}\thanks{E-mail: \url{christopherdavies1234@googlemail.com}}
        \and N. Hamaus\inst{1}
        \and J. J. Mohr\inst{1}
        }

   \institute{University Observatory, LMU Faculty of Physics, Scheinerstr. 1, 81679 Munich, Germany
   \and Department of Physics, Duke University, Durham, NC 27708, USA}

   \date{Received XXX XX, 20XX}

  \abstract
   {Cosmic voids identified in weak-lensing (WL) convergence maps provide a projected probe of underdense regions in the matter distribution. However, the physical connection between these WL voids and voids defined in the 3D large-scale structure remains unclear.}
   {We investigate the correspondence between voids identified in the 3D halo distribution and those detected in WL convergence maps.} 
   {We used 108 realizations of the full-sky lensing simulations, populated with a source galaxy redshift distribution consistent with the Dark Energy Survey Year 3 sample. Catalogs of 3D halo voids and WL voids were constructed using the VIDE and tunnel algorithms, respectively. The spatial association between the two populations was quantified by measuring the angular excess number density of halo voids around WL void centers.}
   {We detected a statistically significant ($S/N \gtrsim 25$) positive correlation between WL voids and low-redshift halo voids ($0 < z_{hv} < 0.5$) at small angular separations, indicating that WL voids trace genuine underdensities in the large-scale matter distribution. The correlation amplitude closely depends on the WL void selection scheme and decreases when stringent peak-amplitude thresholds are applied, thereby reducing the number of detected WL voids and broadening their effective sizes. The signal also displays a strong redshift dependence: halo voids at higher redshift exhibit weaker correlations because of the declining efficiency of the WL kernel, while WL voids identified from higher-redshift source bins produce stronger correlations due to the increased contribution from line-of-sight structures. We additionally examined the impact of halo-void morphology and found that rounder voids with a weaker alignment along the line of sight display marginally stronger associations with WL voids. Our results provide new insights into the contribution of 3D structure to WL-selected underdensities and offer guidance for future observational analyses seeking to interpret WL voids as tracers of the matter field.}
   {}


   \maketitle
   \nolinenumbers

%%%%%%%%%%%%%%%%%%%%%%%%%%%%%%%%%%%%%%%%%%%%%%%%%%%%%%%%%%%%%%
\section{Introduction}
Cosmic voids are vast underdense regions in the large-scale structure (LSS) that can serve as powerful cosmological probes. Their low-density environments make them sensitive to the growth of structure, dark energy, and potential deviations from general relativity \citep{Sheth_2004, Hamaus_2014, Pisani_2019, Davies_2021}.

Galaxies serve as luminous tracers of the underlying dark matter distribution, providing a critical observational link between the visible Universe and its dominant, albeit invisible, mass component. In the standard picture of structure formation, dark matter collapses under gravity to form halos, within which baryons cool and condense to form galaxies \citep{White_1978, Mo_2010}. As a result, the observed galaxy field constitutes a biased sampling of the total matter field, preferentially tracing high-density regions such as filaments and clusters, while avoiding the vast underdense regions that dominate the volume of the cosmic web \citep{Peebles_1980,Kaiser_1984,Cautun2014}. In observations, a variety of techniques have been developed to identify voids using galaxies as tracers. These range from watershed- and tessellation-based methods, such as ZOBOV \citep{Platen_2007, Neyrinck_2008}, to spherical underdensity searches \citep{Colberg_2008, Paz_2023}. In addition, approaches designed to detect circular underdensities in two-dimensional projected galaxy maps have been widely adopted, particularly in weak lensing (WL) studies of cosmic voids \citep{Clampitt_2015, Gruen_2016, Sanchez_2017}. While the precise void boundaries depend on the sampling density, survey geometry, and galaxy bias, statistical ensembles of galaxy-defined voids have been shown to robustly trace large-scale underdensities in both simulations and observations \citep{Cai_2016, Pollina_2017, Fang_2019}. Thus, galaxies provide a practical and robust means of identifying voids, enabling cosmological analyses that leverage the growth and geometry of these underdense regions \citep{Sheth_2004, Hamaus_2016, Pisani_2019, Nadathur_2019, Hamaus_2020, Correa_2021}. Large spectroscopic and photometric surveys such as Sloan Digital Sky Survey (SDSS) and Dark Energy Survey (DES) have made it possible to systematically identify and characterize voids in galaxy distributions \citep{Pan_2012,Hoyle_2012, Sutter_2012, Mao_2017, Sanchez_2017}.

Weak-lensing (WL) convergence maps, reconstructed from the observed shapes of background galaxies in photometric imaging surveys, have already been generated for use in surveys such as DES. In addition to recent developments in detecting 2D voids in the projected galaxy field, it is also possible to identify voids directly in the WL field. Voids identified in this way correspond to deeper line of sight (LOS) underdensities than both 3D and 2D galaxy voids \citep{Davies_2018}. As the WL field is sensitive to the total projected matter distribution, this approach circumvents limitations in our understanding of the galaxy-halo connection. In addition, WL voids have been shown to offer additional information when constraining modified gravity and the constant-\(w\) cold dark matter (wCDM) models \citep{Davies_2019b,Davies_2021}. This approach thereby offers additional valuable avenues to probe underdensities in the LSS. 

Recent studies have begun exploring the 3D structure of underdensities selected in the lensing field \citep{Chang2018,Jeffrey2021,Shimasue_2024}. \citet{Shimasue_2024} investigated the 3D matter distribution around troughs (local minima) identified in the Subaru Hyper Suprime-Cam (HSC) Year 3 (Y3) WL convergence maps \citep{Oguri_2018}. They identified 15 troughs with signal-to-noise ratios (S/Ns) higher than 5.7. They then explored the line-of-sight (LOS) density structure of these troughs utilizing the redshift distributions of photometric luminous red galaxies observed by HSC \citep{Oguri_2018_a} and spectroscopic galaxies detected by Baryon Oscillation Spectroscopic Survey (BOSS) \citep{Reid_2016}. Their results indicate that the lensing signal of most of these troughs can be explained by multiple voids aligned along the LOS, while 2 of the 15 troughs are likely produced by single voids. The authors argue that single voids can indeed generate detectable WL troughs, provided they are nonspherical and significantly elongated along the LOS.

In this work, we measure the correlation between 3D voids identified in the halo distribution and WL voids identified in 2D WL convergence maps. Quantifying this connection will offer further insights into the physical nature of WL voids and the relations between their characteristic scales and properties. All analyses presented in this work have been performed on mock halo catalogs and simulated WL convergence maps derived from the \citet{Takahashi_2017} simulations (hereafter T17).

This paper is organized as follows. Section~\ref{sect:data} describes the simulation data we used in this work. The methods for identifying voids in the halo distribution and from WL convergence maps are detailed in Section~\ref{sect:void_finders}. The methodology for measuring the void cross-correlations is described in Section~\ref{sect:cross-correlation}, where we also present the measurement results and corresponding interpretations on the mock data. We conclude in Section~\ref{sect:conclusion} with a summary of our findings.
%%%%%%%%%%%%%%%%%%%%%%%%%%%%%%%%%%%%%%%%%%%%%%%%%%%%%%%%%%%%%%

\section{Simulation data}
\label{sect:data}

\begin{figure}
    \centering
    \includegraphics[width=\columnwidth]{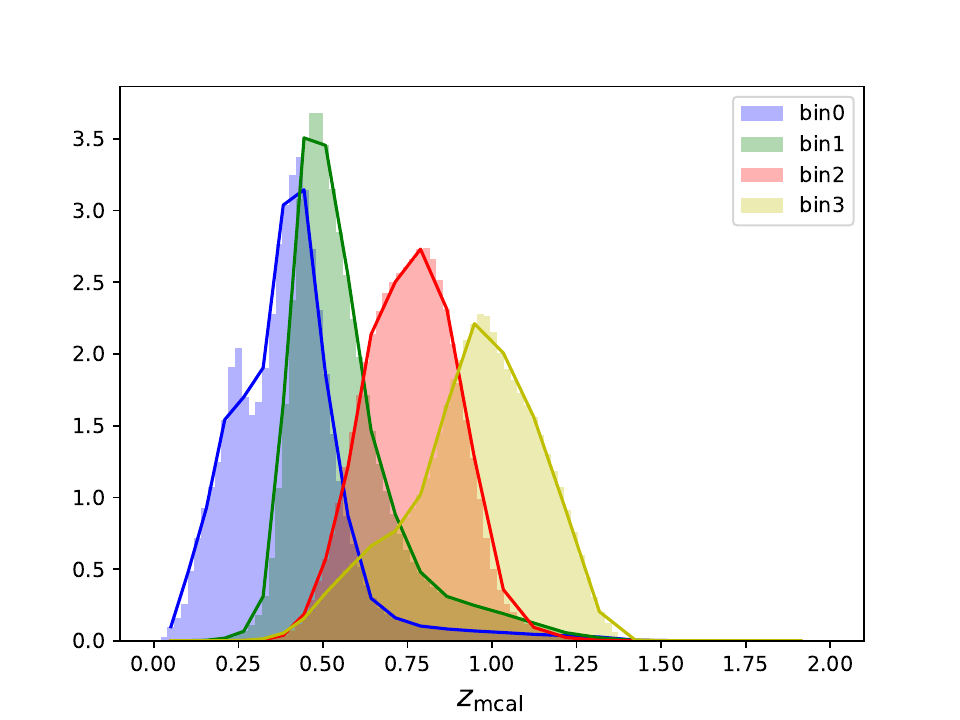}
    \caption{DNF-estimated photometric redshift distributions for galaxies in the DES Year 3 shape catalog across four tomographic bins.}
    \label{fig:nz_mcal}
\end{figure}

\label{sect:mock_data}
 T17 simulations are a suite of 108 full‑sky gravitational lensing simulations, created using high‑resolution cosmological N‑body runs with cosmological parameters consistent with WMAP 9yr results \citep{Hinshaw_2013}, and combined with multiple‑lens‑plane ray‑tracing. The simulations were generated using a system of nested cubic boxes designed to reproduce the large-scale mass distribution of the Universe, with side lengths $L, 2L, 3L, ...$ and $L=450 h^{-1}\rm{Mpc}$. Each box, containing $2048^3$ particles, was replicated eight times around a fixed vertex corresponding to the observer's position under periodic boundary conditions. The simulations produced convergence and shear maps spanning redshifts from $z\approx0.05$ to $5.3$, with radial slices spaced every $150h^{-1}\rm{Mpc}$ comoving distance. Shear and convergence field maps are provided in the form of \texttt{HEALPIX} \citep{Healpix_ref} maps with resolution of \texttt{NSIDE}$=4096\ \rm{and}\ 8192$. We used maps with \texttt{NSIDE} $=4096$, corresponding to a pixel angular size of 0.86 arcminutes and providing a sufficient resolution for identifying voids in the convergence fields. These maps also provided the corresponding halo catalogs containing positions, redshifts, velocities, and masses.

 The original T17 convergence maps are available in the form of discrete source planes. We constructed DES-like effective convergence maps by LOS weighting these maps. To ensure the mock convergence planes reflected the realistic redshift distributions (i.e., as seen in observations), we adopted the redshift distributions from the Dark Energy Survey Year 3 (DES Y3) WL source catalog \citep{Gatti_2021, DESY3_photometric} as a reference for population. DES Y3 covers the first three years (2013-2016) of DES observations with the Dark Energy Camera on the 4-meter Blanco Telescope at CTIO, providing deep grizY imaging over $\sim5,000 \rm{deg}^2$ to a limiting magnitude of $r\sim24(AB)$ for extended sources \citep{DES_2016}. Photometric redshifts for Y3 GOLD objects were estimated using BPZ \citep{BPZ_2000}, DNF \citep{De_Vicente2016}, and ANNz2 \citep{Sadeh_2016}. We adopted the mean photo-z values from the DNF algorithm, which infers redshifts via local hyperplane fits in color-magnitude space. As illustrated in Fig.~\ref{fig:nz_mcal}, each source galaxy was assigned to one of four redshift bins, based on its combined photometric and clustering-informed likelihood.

 For each of the 108 T17 realizations, we constructed a tracer 
 halo catalog and produced five convergence maps: four corresponding to the DES tomographic redshift bins and one based on the full redshift distribution. The tracer halo catalog is built by selecting only parent halos (i.e., those with \texttt{PID}$\neq -1$) having virial masses greater than $10^{13} M_\odot h^{-1}$. 
 
The observed convergence field $\kappa(\pmb {\theta})$ measured from a population of source galaxies is given by the weighted projection of the full convergence as a function of comoving distance $\kappa(\pmb{\theta},\chi)$, where the weights are given by the source galaxy distribution in comoving distance \citep{Kilbinger_2014}. To more closely resemble observational data using the T17 convergence maps, we applied weights to each map based on the point-estimated redshift distribution of galaxies from the DES Y3 WL source catalog at the corresponding redshift,

\begin{equation}
    \kappa(\pmb {\theta}) = \int_{0}^{z}n(z')\kappa(\pmb {\theta}, z') \,dz' = \Sigma_{i} n(z_i)\kappa(\pmb {\theta}, z_i) \, ,
    \label{eq:kappa_map_construction}
\end{equation}

where $n(z_i)$ denotes the normalized number density of source galaxies in the DES Y3 WL source catalog for a given redshift bin. The T17 convergence maps naturally include the non-Gaussian structures generated by nonlinear gravitational clustering through full ray-tracing simulations. However, we note that our analysis did not additionally incorporate observational shear calibration bias, intrinsic ellipticity noise, or instrumental systematics.

%%%%%%%%%%%%%%%%%%%%%%%%%%%%%%%%%%%%%%%%%%%%%%%%%%%%%%%%%%%%%%

\section{Void finders}
\label{sect:void_finders}

\subsection{Halo voids}
\label{sect:tracer_voids}

To identify cosmic voids in the halo distribution, we used the publicly available Void IDentification and Examination (VIDE) toolkit \citep{sutter_2015}, which serves as a wrapper around an enhanced version of Zones Bordering On Voidness (ZOBOV; \citet{Neyrinck_2008}). While ZOBOV was originally designed for void finding in simulations with periodic boundary conditions, VIDE extends its capabilities to accommodate observational data, incorporating features such as boundary handling and mask support. The void-finding process can be summarized as follows:
\begin{enumerate}
    \item Voronoi tessellation: A 3D Voronoi tessellation is applied to the tracer catalog. Each tracer is assigned a Voronoi cell, which defines the region closer to it than to any other particle. The local density is then estimated as the inverse of the cell’s volume. 
    \item Identification of density minima: Local density minima are identified by locating tracer particles whose Voronoi cells are larger than those of all their neighbors.
    \item Zone construction: Starting from each density minimum, adjacent cells with monotonically increasing density are iteratively grouped together. This forms zones that correspond to local underdense basins in the density field.
    \item Merging: A watershed transform \citep{Platen_2007} is used to merge adjacent zones into larger voids, building a hierarchical structure of voids and subvoids. 
    \item Void characterization: The above zones correspond to the identified voids. Each void is assigned an effective radius, $R_v$, defined as the radius of a sphere with the same total volume. The void center is computed as the volume-weighted barycenter of all contributing Voronoi cells.
\end{enumerate}

\begin{figure}
    \centering
    \includegraphics[width=1.\linewidth]{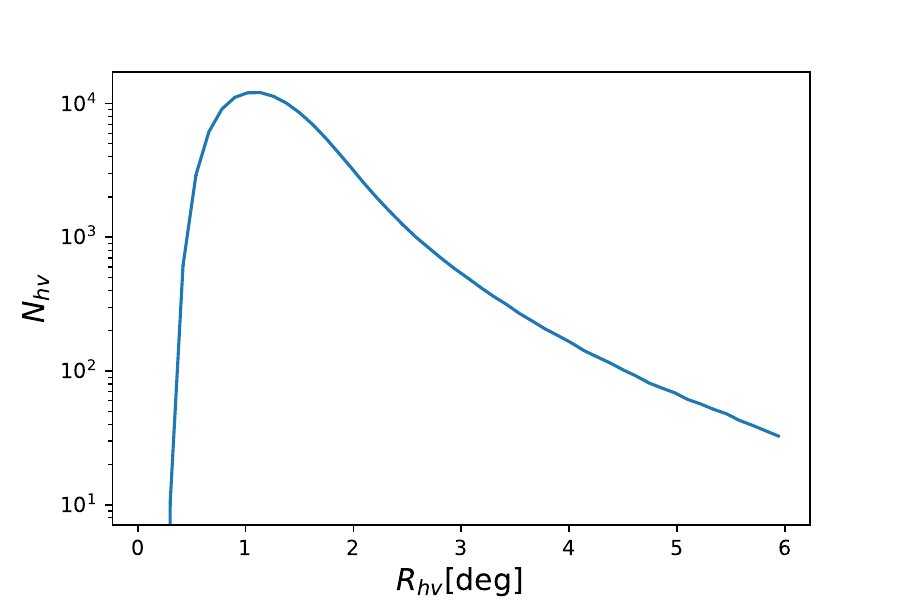}
    \caption{Mean 3D halo void counts as a function of effective angular radius, $R_{hv}$, averaged over 108 mock realizations. The voids are grouped into 50 linearly spaced radius bins. Error bars indicate the standard error of the mean in each bin, but are mostly smaller than the line width and therefore not visible in the figure.}
    \label{fig:size_func_hv}
\end{figure}

We ran the VIDE void finder on the T17 simulated halo catalogs described in Sec.~\ref{sect:mock_data}, using parent halos with $M_{\rm{vir}} > 10^{13}M_{\odot}h^{-1}$ as tracers, for all 108 realizations, resulting in 108 corresponding 3D halo void catalogs. Fig.~\ref{fig:size_func_hv} shows the mean number of halo voids identified in 50 linearly spaced angular-radius bins, averaged over the 108 mock realizations. Each catalog was further subdivided into three redshift bins: $0 < z_{hv} < 0.5$, $0.5 < z_{hv} < 1.0$, and $1.0 < z_{hv} < 1.5$. This allowed us to investigate the redshift dependence of their correlations with WL voids.

\subsection{WL voids}

The tunnel algorithm \citep{Cautun_2018} is a 2D void finder that identifies the largest circles that are empty of tracers \footnote{The tunnel algorithm used in this work is publicly available at \href{https://github.com/chrisdavies234/tunnel_finder}{https://github.com/chrisdavies234/tunnel\_finder}}. This is achieved by first constructing a Delaunay triangulation out of a set of discrete points \footnote{Note that the Delaunay triangulation corresponds to the dual graph of the Voronoi diagram. The Voronoi diagram is a popular method for estimating the underlying density field sampled by a set of tracers, and commonly employed in other void finding algorithms.}. Each cell in the Delaunay triangulation is defined as a triangle whose vertices correspond to a tracer, where each triangle does not enclose any tracers. These cells are then used to define a circumcircle whose circumference intersects the three vertices of its corresponding triangle and, hence, the circumcircles also contain no tracers. These circumcircles correspond to the tunnels (2D voids) identified by the algorithm. 

In this work, we use the positions of WL peaks, defined as local maxima in the smoothed convergence field, as tracers for the tunnel algorithm. Peaks are identified at the HEALPIX level by selecting pixels whose convergence values exceed those of all neighboring pixels. Prior to peak identification, the convergence maps are smoothed with a Gaussian kernel of a width expressed by $\sigma_{\rm{smooth}}$ in order to suppress fluctuations. We considered three different WL peak-selection schemes:
\begin{itemize}
    \item No threshold: all identified local maxima are retained, regardless of the sign or amplitude of the convergence value;
    \item Positive peaks only ($\kappa > 0$): only local maxima with positive convergence values are retained;
    \item High-threshold peaks ($\kappa > 0.01$): only local maxima with convergence amplitudes greater than 0.01 are retained.
\end{itemize}We also note that the algorithm used here operates directly on the curved sky, performing the Delaunay triangulation directly on the surface of a unit sphere, where previous studies with the tunnel algorithm have employed the flat sky approximation \citep[e.g.][]{Cautun_2018, Davies_2018}. After constructing all possible circumcircles, we removed those associated with Delaunay triangles whose minimum internal angle was smaller than 20 degrees. Such triangles are highly elongated and occupy only a small fraction of the area enclosed by their corresponding circumcircles. As a result, the associated circumcircles do not accurately trace compact underdense regions and might artificially inflate the inferred WL void size.

Next, we removed circumcircles whose centers are enclosed within a larger circumcircle, which gives priority to identifying the largest objects, while simultaneously reducing the overlap of the circumcircles. This minimizes the duplicate information throughout the catalog.   

\begin{figure}
    \centering
    \includegraphics[width=1.\linewidth]{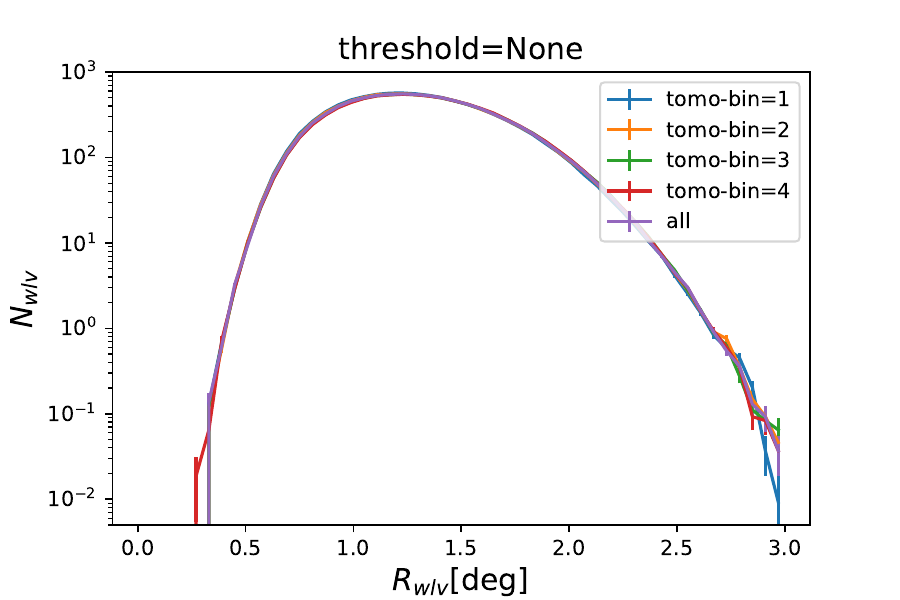}
    \includegraphics[width=1.\linewidth]{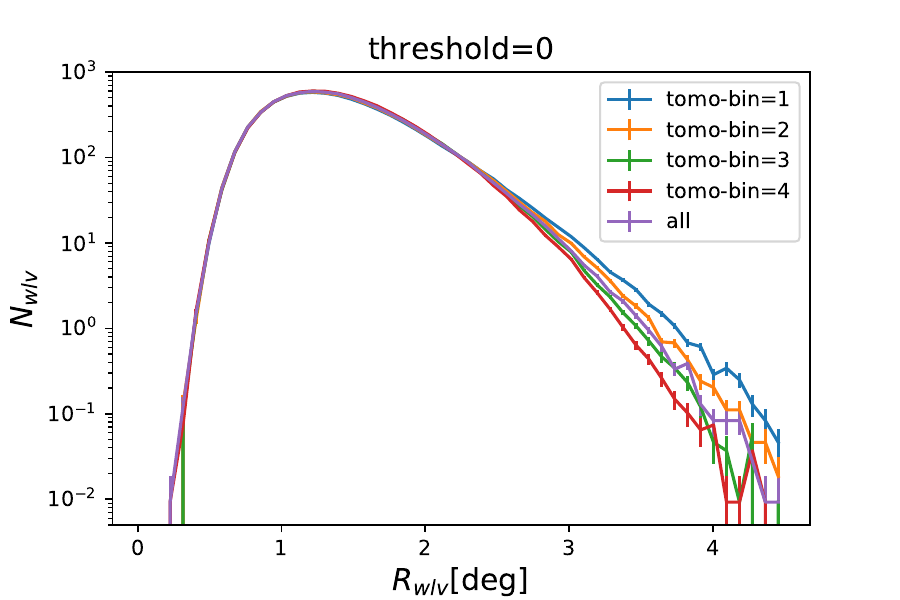}
    \includegraphics[width=1.\linewidth]{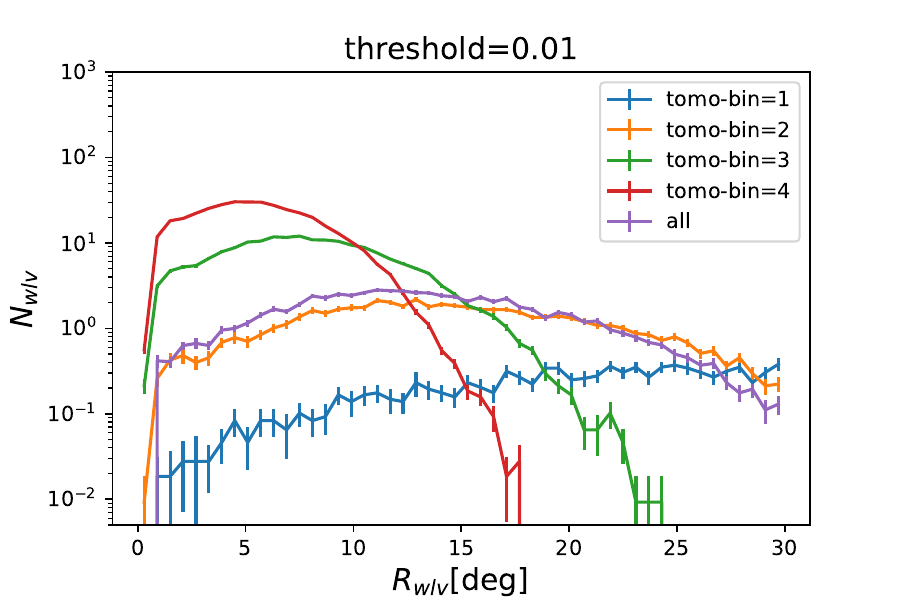}
    \caption{Using 108 mock realizations, we compute the mean void size functions from WL maps constructed based on the redshift distributions of different DES-Y3 tomographic bins. Void detection was performed under three conditions: without applying any selection on the peaks (top), using only positive peaks (middle), and selecting peaks with values greater than 0.01 (bottom).}
    \label{fig:size_func_tomo_bins}
\end{figure}

We applied the WL void finder to all five convergence maps described in Sec.~\ref{sect:mock_data} for each of the 108 realizations. WL void detection is performed on three different tracer catalogs: the full WL peak catalog without any additional cuts, WL peaks with only positive amplitudes, and WL peaks with amplitudes above a threshold of $\kappa=0.01$. The resulting mean void radius distributions (the void size functions) are presented in Fig.~\ref{fig:size_func_tomo_bins}. In the unfiltered case (top panel), there is no visible redshift dependence in the void size distribution. Once a positive peak threshold is applied (middle panel), a clear redshift trend begins to emerge: the number of large voids decreases at higher redshift. This effect becomes even stronger when applying a stricter threshold, such as $\kappa>0.01$ (bottom panel), where high-redshift bins show noticeably smaller voids. Voids identified in the full, nontomographic convergence map (“all”) fall near the midpoint of this redshift trend, consistent with their effective mean redshift when all bins are combined. This can be seen in both of the bottom two panels. 

The impact of the peak-amplitude cuts on the tracer population is substantial. Relative to the unfiltered peak catalog, the $\kappa > 0$ selection retains approximately $85\%$ of all identified WL peaks, while the more stringent $\kappa > 0.01$ threshold retains no more than $10\%$. Consequently, the latter selection dramatically reduces the surface density of tracers available to the tunnel finder, leading to the identification of fewer and larger WL voids. The reduction in tracer density naturally increases the typical separation between neighboring peaks. Since the tunnel algorithm identifies voids as empty regions enclosed by surrounding peaks, this leads to larger characteristic WL-void radii and a smaller total number of detected voids.

This redshift dependence arises from the evolving WL peak distribution. Although the total number of WL peaks remains roughly constant with redshift, the fraction of high-amplitude peaks increases as the source redshift rises. \cite{Ferlito_2025} argued that at higher redshift, the integrated mass distribution produces stronger lensing signals, broadening the peak amplitude distribution. Consequently, the number of peaks above a fixed signal threshold grows with redshift, increasing the surface density of identified peaks. Since the tunnel finder defines voids as regions enclosed by surrounding peaks, an increase in peak density naturally results in smaller identified voids. This behavior is analogous to what occurs in 3D halo-based void finding when the halo mass threshold is lowered, leading to a denser tracer population and correspondingly smaller voids.

\begin{figure}
    \centering
    \includegraphics[width=1.\linewidth]{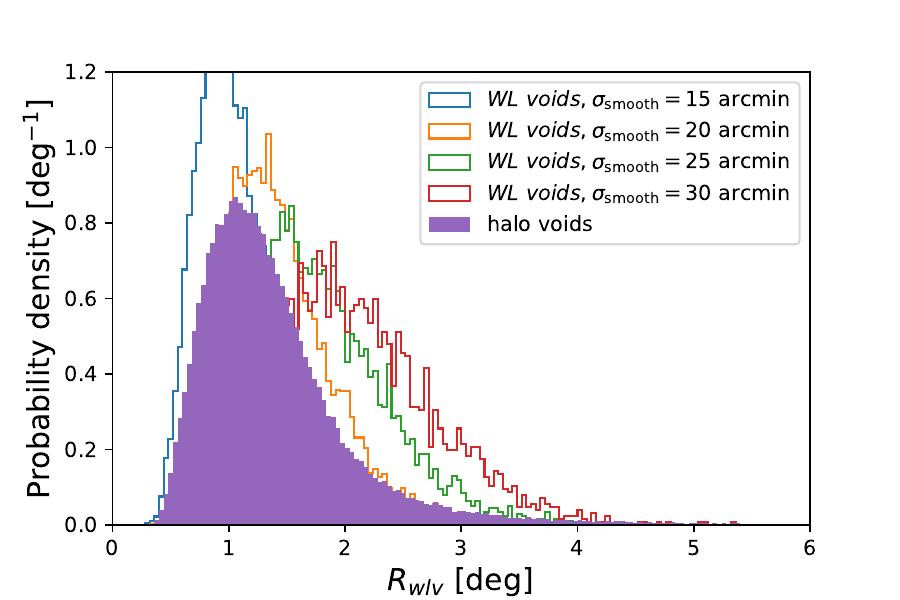}
    \caption{Impact of convergence map smoothing on WL void sizes. The distributions are computed using 200 linearly spaced bins in $R_{wlv}$ and normalized to probability density. Varying the smoothing scale strongly alters the WL void size distribution: larger smoothing suppresses small-scale features, producing fewer and larger voids, while smaller smoothing retains finer structure and yields a broader distribution of smaller voids. The size distribution of the 3D halo voids is shown in purple for comparison.}
    \label{fig:smooth_scale}
\end{figure}

As shown in Fig.~\ref{fig:smooth_scale}, the choice of smoothing scale applied to the convergence map has a substantial impact on the resulting WL void size function. Increasing the smoothing scale suppresses small-scale features and merges nearby peaks, leading to the identification of larger and fewer WL voids, whereas smaller smoothing scales retain finer structures and yield a broader distribution of smaller voids. To facilitate a more meaningful assessment of the projected spatial correspondence between WL voids and 3D halo voids, we adopted a smoothing scale of $\sigma_{\rm{smooth}} = 20$ arcmin. At this scale, the WL void size function most closely matches that of the 3D halo voids, minimizing discrepancies that arise purely from differences in tracer density or smoothing methodology.

Table~\ref{tab:void_counts} summarizes the typical sizes of the halo void and WL void catalogs used throughout this work. The reported values correspond to the mean numbers of identified voids per realization across the 108 T17 simulations. The aggressive $\kappa > 0.01$ selection substantially reduces the WL void sample size.

\begin{table}
\centering
\caption{Mean numbers of identified voids}
\label{tab:void_counts}
\begin{tabular}{lcc}
\hline
Sample & Selection & Number of voids per realization \\
\hline
Halo voids & --- & $N_{\rm hv} = 118701 \pm 399$ \\
WL voids & No threshold & $N_{\rm wlv} = 8382 \pm 93$ \\
WL voids & $\kappa > 0$ & $N_{\rm wlv} = 6924 \pm 90$ \\
WL voids & $\kappa > 0.01$ & $N_{\rm wlv} = 141 \pm 130$ \\
\hline
\end{tabular}
\tablefoot{Values are averaged over the 108 T17 mock realizations, with uncertainties representing the realization-to-realization standard deviation.}
\end{table}

%%%%%%%%%%%%%%%%%%%%%%%%%%%%%%%%%%%%%%%%%%%%%%%%%%%%%%%%%%%%%%
\section{Void cross-correlation}
\label{sect:cross-correlation}

To investigate the relationship between voids identified in the halo field and those detected in WL convergence maps, we measured the projected angular number density contrast profile of halo voids around the positions of WL void centers. This measurement provides a quantitative way to assess whether the two types of voids preferentially occupy similar regions of the sky and how their characteristic scales are related.

\subsection{Methodology}
For each WL void, we computed the surface densities of halo defined voids by counting the number of 3D halo void centers $N(\pmb {\theta})$ falling within an angular bin $[\pmb {\theta} - \delta\pmb {\theta}/2, \pmb {\theta}+\delta\pmb {\theta}/2]$. We compare the result to their mean surface number density over the full sky. The average surface density of halo voids around WL voids is given by
\begin{equation}
    n_{hv}(\pmb {\theta})=\frac{1}{N_{wlv}}\sum\limits_{i=1}^{N_{wlv}}\frac{N_i(\pmb {\theta})}{A(\pmb {\theta})},
\end{equation}
where $N_{\rm{wlv}}$ is the total number of WL voids considered, $N_i(\pmb {\theta})$ is the number of 3D halo voids found within the angular bin around the i-th WL void, and $A(\pmb {\theta})$ is the area of the annular bin. The contrast profile is then
\begin{equation}
    \delta_{whv}(\pmb {\theta}) = \frac{n_{hv}(\pmb {\theta})}{\overline{n}_{hv}}-1,
\end{equation}
where $\overline{n}_{hv}$ is the mean surface number density of halo voids computed over the full sky area.

This stacked profile was computed for each of the 108 T17 realizations, which we used to calculate the mean signal and associated error bars while robustly accounting for sample variance. The final result is obtained by averaging the individual profiles over all realizations. The elements in the covariance matrix was estimated via
\begin{equation}
    \matr{C}_{mn}=\frac{1}{N_{\rm{r}}-1}\sum\limits_{i=1}^{N_{\rm{r}}}(\delta_{whv}^{i,m}-\bar{\delta}^{m}_{whv})(\delta_{whv}^{i,n}-\bar{\delta}^{n}_{whv})^T,
\end{equation}
where $N_{\rm{r}}=108$ is the number of realizations, $i$ the realization index, $(m,n)$ the radial-bin indices, and $\bar{\delta}_{whv}$ is the mean profile. The error bars of the profile correspond to the standard error on the mean measurement: $\sigma_i=\sqrt{\matr{C}_{ii}/N_{\rm{r}}}$. Since $N_{\rm{r}}$ is finite, the inverse of the sample covariance, $\matr{C}^{-1}$, provides a biased estimate of the true matrix because the inversion amplifies noise in the limited sample. To correct for this bias, we applied the Hartlap correction \citep{Hartlap_2007}, which rescales the inverse covariance by a factor of
\begin{equation}
    \alpha=\frac{N_{\rm{r}} - N_{\rm{d}} -2}{N_{\rm{r}}-1},
\end{equation}
where $N_{\rm{d}}=10$ is the dimension of the data vector. Thus, the significance of the mean profile can be calculated by
\begin{equation}
    \chi^2=\bar{\delta}_{whv}^T(\matr{C}/N_{\rm{r}})^{-1}\bar{\delta}_{whv}.
\end{equation}
For ease of interpretation, we report the corresponding detection significance,
\begin{equation}
    \rm{S/N}=\sqrt{\chi^2},
\end{equation}
which quantifies the significance of rejecting the null hypothesis of no correlation.

\subsection{Measurements on mocks}
\label{sect:mock_measurements}
We carried out this measurement for WL void catalogs identified under three peak-selection schemes: no additional cuts, positive peaks only, and peaks with $\kappa>0.01$. The analysis was repeated for each mock DES Y3 tomographic bin and the combined source redshift sample, for each halo void redshift bin, and for the combined halo void sample spanning $0 < z_{hv}<1.5$.

\begin{figure*}
    \centering
    \includegraphics[width=1.\linewidth]{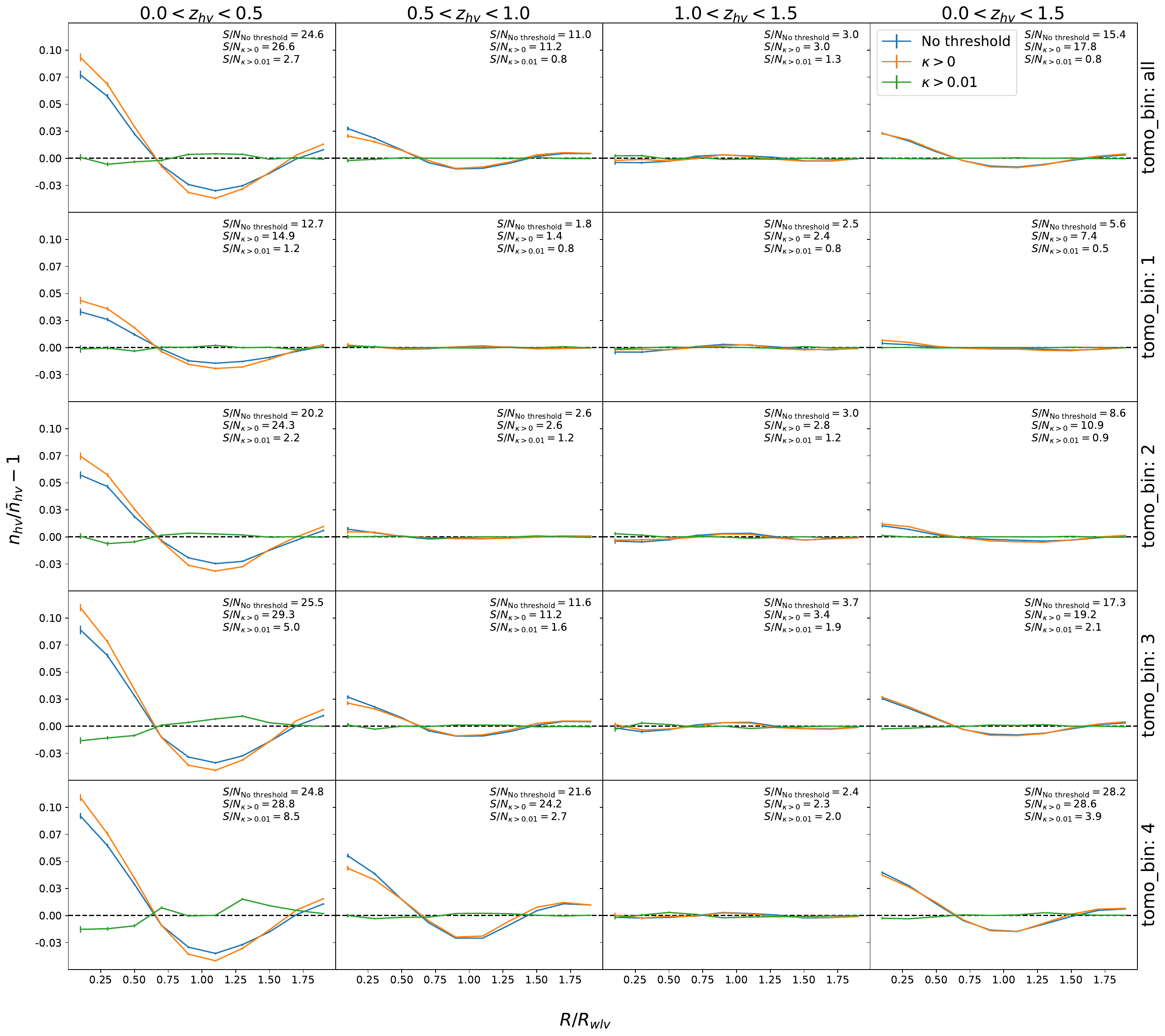}
    \caption{Mean excess projected number density profiles, $\delta_{whv}(\pmb {\theta})$, of halo voids around WL voids, 
    shown together with their associated standard errors and corresponding S/N (relative to the null hypothesis of no correlation), averaged over 108 realizations. Each row corresponds to a different WL source redshift selection from top to bottom: full DES Y3 source sample, tomographic bin 1, bin 2, bin 3, and bin 4. Meanwhile, each column corresponds to a different 3D halo-void redshift bin (from left to right: $0 < z_{hv} < 0.5$, $0.5 < z_{hv} < 1.0$, $1.0 < z_{hv} < 1.5$ and the combined sample $0 < z_{hv} < 1.5$). Within each panel, different colored curves show results for WL voids identified under three peak-selection schemes: no threshold, positive peaks only, and peaks with $\kappa>0.01$. For the lowest-redshift 3D halo-void bin ($0 < z_{hv} < 0.5$), the profiles reveal a clear positive correlation at small angular separations ($\theta \lesssim R_{\rm{wv}}$), indicating that low-redshift 3D halo voids are preferentially located near WL void centers in projection.}
    \label{fig:profile}
\end{figure*}

Fig.~\ref{fig:profile} presents the mean cross-correlation profiles and the corresponding significances between WL voids and halo voids. For the lowest redshift bin of 3D halo voids, the results reveal a positive correlation at small angular separations ($\theta \lesssim R_{\rm{wv}}$), confirming that in 2D approaches, 3D halo voids are preferentially located near WL void centers. The amplitude of the correlation varies systematically with the WL void peak-selection scheme:
\begin{itemize}
    \item For the lowest-redshift halo void sample ($0<z_{hv}<0.5$) and the WL void sample constructed from the full source redshift distribution, both the no threshold and $\kappa > 0$ selections yield highly significant detections, with S/Ns of $24.6$ and $26.6$, respectively. The $\kappa > 0$ selection produces the strongest signal, indicating a slightly higher degree of spatial correspondence between WL voids and halo voids when only positive WL peaks are retained.
    \item In contrast, the $\kappa > 0.01$ selection reduces the detection significance to $2.7$, reflecting the substantially smaller WL void sample size and the larger characteristic void radius produced by this more stringent peak threshold.
\end{itemize}

The correlation strength shows a strong dependence on halo redshift. In the lowest redshift 3D halo void bin ($0<z_{hv}<0.5$), the signal is the strongest, while it decreases progressively in the intermediate ($0.5<z_{hv}<1.0$) and high  ($1.0<z_{hv}<1.5$) bins. This trend is consistent with the physics underpinning the lensing measurements. The lensing efficiency peaks at half the comoving distance between the source and observer; for the DES Y3 source redshift distribution, this occurs in the lowest 3D halo void redshift bin. We also note that part of this trend may be influenced by the increasing effective projection length associated with higher redshift bins. Higher redshift source distributions generally span broader ranges in comoving distance, which increases the mixing of structures along the LOS and can dilute the spatial correspondence between projected WL voids and individual 3D halo voids. Consequently, the observed redshift dependence likely reflects a combination of lensing kernel efficiency and redshift projection effects. Disentangling these contributions in detail would require a more controlled tomographic analysis and is left for future works.

The amplitude of the cross-correlation between WL voids and 3D halo voids also exhibits a clear source redshift dependence, increasing for higher-redshift tomographic WL convergence maps. This trend arises because higher-redshift source planes integrate over a longer LOS, thereby accumulating more large-scale structure. The enhanced projection of mass fluctuations strengthens the WL convergence signal and, consequently, amplifies the correlation between the projected and 3D halo void populations.

The overall shape of the cross-correlation function provides further physical insight into the spatial relationship between WL and 3D halo voids. The signal is positive at small separations within the interior of the WL void, indicating an increased probability of finding a 3D halo void along the same LOS. This alignment suggests that WL voids trace genuine underdensities in the matter field. At the void boundary, the correlation becomes negative, reflecting a deficit of 3D halo voids in these regions. This anti-correlation occurs because the WL void edges coincide with projected overdense ridges populated by WL peaks. Beyond the void boundary, the correlation signal asymptotically approaches zero as the spatial separation increases, indicating that regions far from the WL void centers are statistically uncorrelated with them and represent the background level.

An interesting feature arises in the high-threshold $\kappa>0.01$ case, where some tomographic combinations exhibit mildly negative cross-correlation amplitudes. In most cases, however, the corresponding $S/N$ values are low, suggesting that these anti-correlations may simply reflect statistical noise due to limited WL void sample size or weak S/N. 
One exception is the combination of low-redshift halo voids and high-redshift WL convergence maps, where the negative signal is more pronounced. A possible interpretation is that the stringent peak threshold preferentially selects very large WL voids whose projected areas include substantial LOS structures, thereby weakening the correspondence with genuine 3D underdensities. In this regime, projection effects and residual overdense structures may partially wash out the expected positive association between WL and halo voids. However, additional investigation will be required to robustly determine the physical origin of this behavior.

\subsection{Splitting properties on halo voids}
To better understand the factors driving the observed spatial correlation between WL voids and halo voids, we performed a series of targeted analyses on various subsamples of the 3D halo void population. In this test, we focused on WL voids identified in the convergence map constructed from the full DES Y3 redshift distribution, which provides a representative high-S/N sample. Only positive convergence peaks are used in the WL void identification. The halo void catalogs are then divided according to three properties: redshift, ellipticity, and elongation with respect to the LOS:

\begin{enumerate}
    \item Redshift, $z_{hv}$: We take a close look at the first 2 redshift intervals defined in \ref{sect:tracer_voids}.
    \item Ellipticity, $e_{hv}$: Ellipticity values are given by VIDE via estimating from the inertia tensor of the member halos. Smaller values of $e_{hv}$ indicate more isotropic, approximately round void shapes, while larger values correspond to more elongated or irregular voids. The sample is split into “low ellipticity” and “high ellipticity” subsamples at $e_{hv}=0.15$, enabling an assessment of whether more elongated or more spherical halo voids correlate differently with WL voids.
    \item Orientation relative to the LOS: Using the principal axes derived from the inertia tensor, we measure the cosine of alignment angle $|cos(\Theta)|$, along with the corresponding ellipticity, we define two subsamples: the more aligned and nonspherical one with $|cos(\Theta)|\geq0.7\ \rm{and}\ e_{hv}\geq0.2$, and the less aligned and spherical one with $|cos(\Theta)|\leq 0.3\ \rm{and}\ e_{hv}\leq0.15$. Halo voids whose major axes are more closely aligned with the LOS are expected to be more effective WL lenses due to enhanced projected underdensity, as suggested in earlier works \citep{Shimasue_2024}.
\end{enumerate}

\begin{figure}
    \centering
    \includegraphics[width=1.\linewidth]{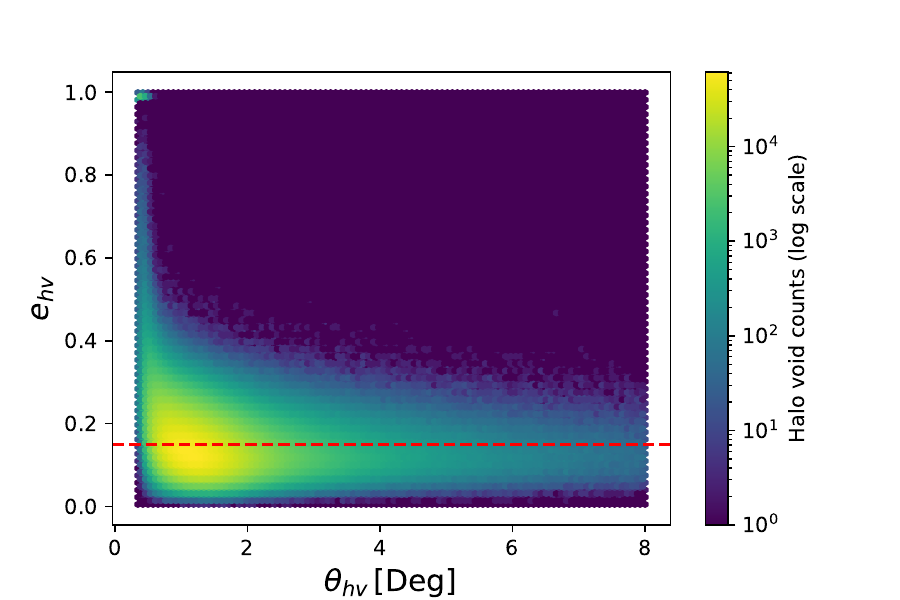}
    \includegraphics[width=1.\linewidth]{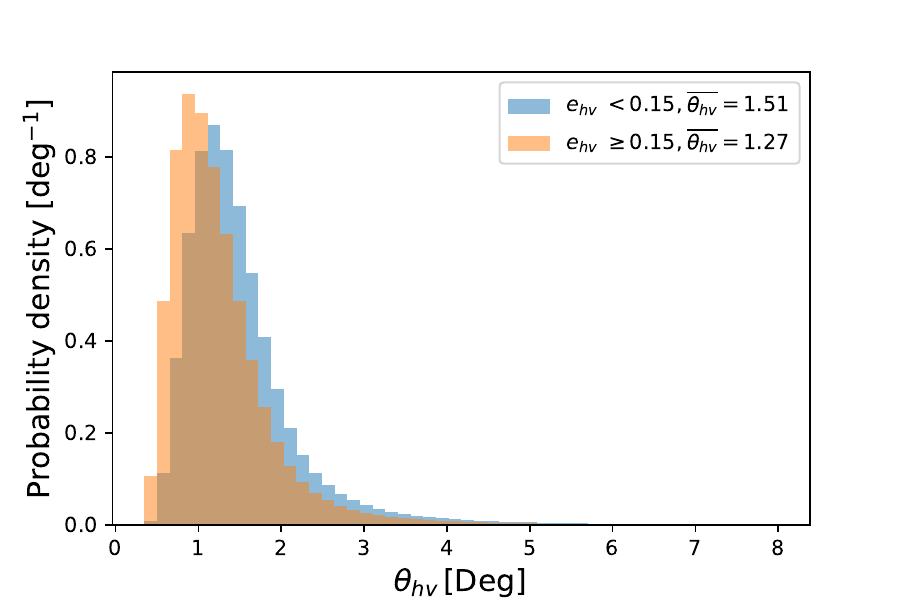}
    \caption{Hexagonal-binned density map of halo void ellipticity as a function of an angular size, $\theta_{hv}$ (top panel). The color scale shows the number of halo voids per hexagonal bin on a logarithmic scale, combining all halo voids identified across the 108 realizations. The red dashed line marks the ellipticity threshold used to divide the sample. We also plotted the distribution of angular size, $\theta_{hv}$, of the two subsamples (bottom panel). Results indicate that highly elliptical voids tend to be smaller, which reduces their influence on the WL field.}
    \label{fig:ellipticity_distribution}
\end{figure}

\begin{figure}
    \centering
    \includegraphics[width=1.\linewidth]{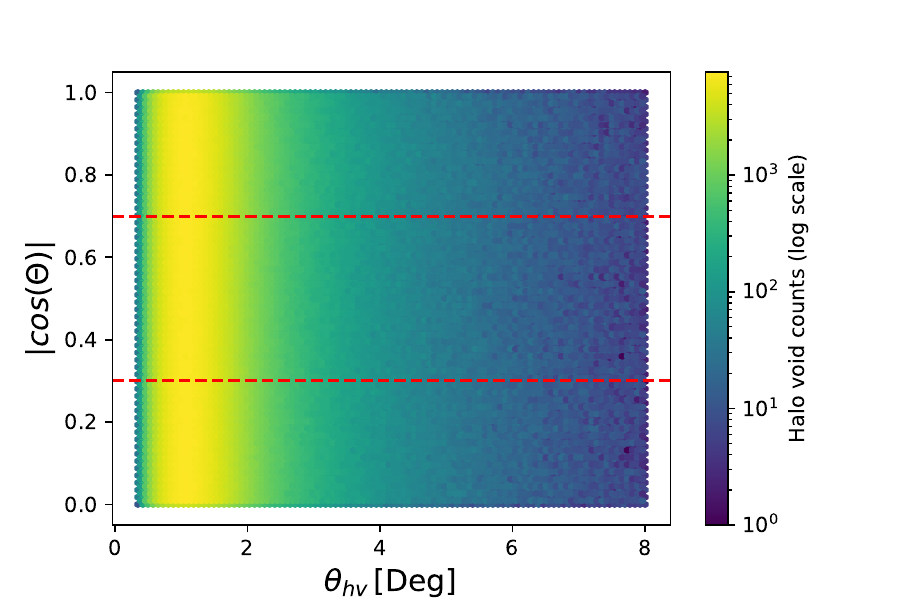}
    \caption{Hexagonal-binned density map showing the cosine of the angle between the principal axes of halo voids and the LOS to their centers, shown as a function of angular size, $\theta_{hv}$. The color scale indicates the number of halo voids per hexagonal bin on a logarithmic scale, using all halo voids identified across the 108 T17 realizations.}
    \label{fig:alignment_distribution}
\end{figure}

To illustrate the properties of these subsamples, Fig.~\ref{fig:ellipticity_distribution} shows the distribution of halo void ellipticity as a function of angular radius, together with the radius distributions of the low- and high-ellipticity samples. Fig.~\ref{fig:alignment_distribution} shows the LOS alignment parameter $|cos(\Theta)|$ as a function of halo void angular radius. 

\begin{figure}
    \centering
    \includegraphics[width=1.\linewidth]{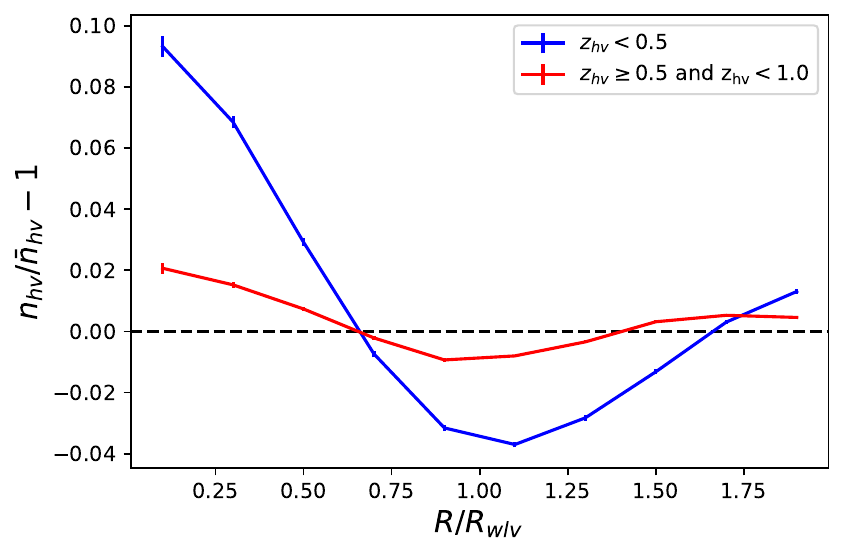}
    \includegraphics[width=1.\linewidth]{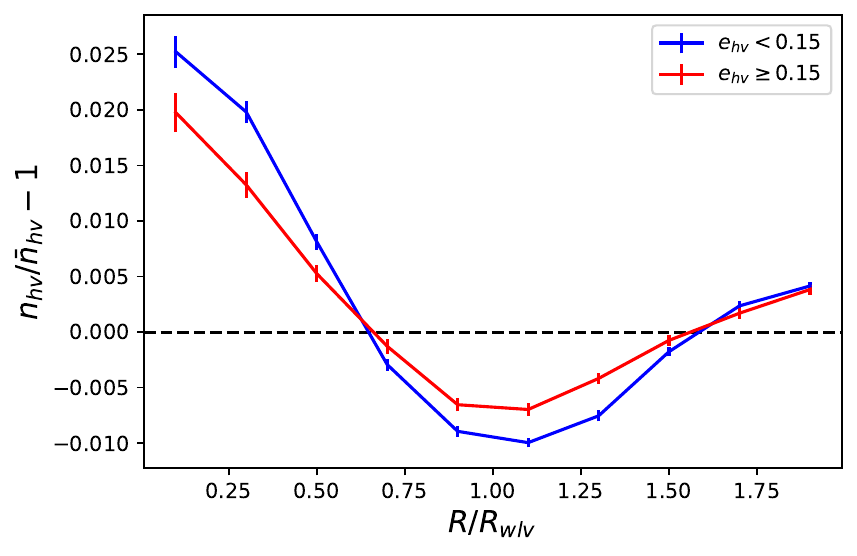}
    \includegraphics[width=1.\linewidth]{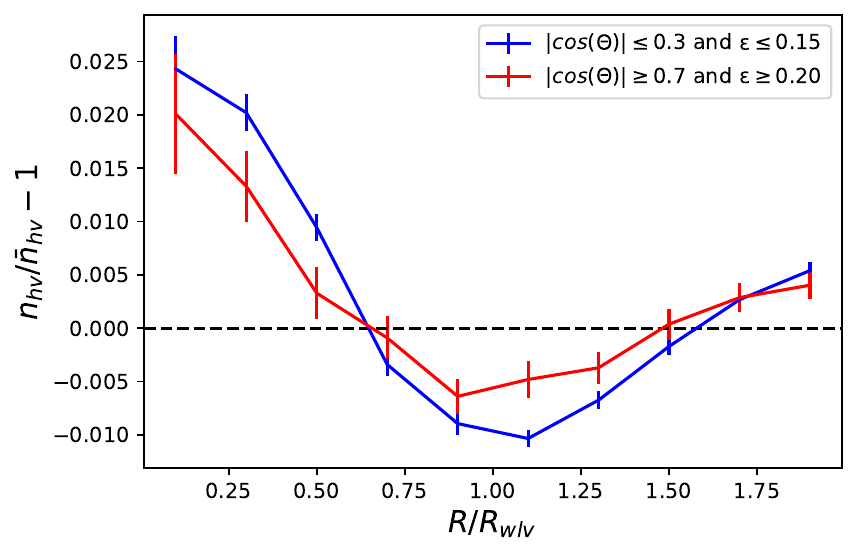}
    \caption{Radial profiles of average excess projected number density of halo voids centered on WL voids, along with the split by the redshifts of halo voids (top), the ellipticities of halo voids (middle) and the elongation along the LOS of halo voids (bottom). Error bars denote the standard error on the mean estimated from the 108 T17 realizations.}
    \label{fig:split_tracer_voids_properties}
\end{figure}

\begin{figure}
    \centering
    \includegraphics[width=1.\linewidth]{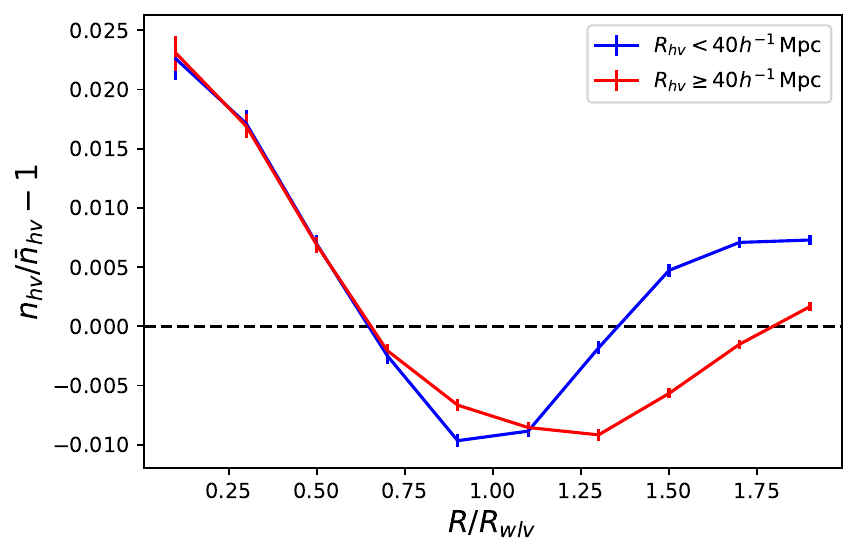}
    \caption{Radial profiles of average excess projected number density of halo voids centered on WL voids, split at $R_{hv}= 40 h^{-1}\,\mathrm{Mpc}$. Error bars denote the standard error on the mean estimated from the 108 T17 realizations.}
    \label{fig:split_tracer_voids_radius}
\end{figure}

For each subsample, we measure the average (over all 108 realizations of T17 simulation) excess projected number density profile of halo voids around the corresponding WL voids, following the methodology in Section 3.1. The results are shown in Fig.~\ref{fig:split_tracer_voids_properties}. They indicate:
\begin{itemize}
    \item The low-redshift subsample consistently yields the strongest positive correlation, in agreement with the results in Sect.~\ref{sect:mock_measurements}. This trend is consistent with expectations from lensing theory: the DES Y3 sources have redshift distribution which peaks at around 0.5 (Fig.~\ref{fig:nz_mcal}). At low redshift, halo voids lie closer to the peak sensitivity of the WL kernel. At higher redshift, the lensing kernel becomes broader and the underdensities are more easily diluted due to integration of the density field along the LOS.
    \item Less elliptical voids show a marginally enhanced correlation amplitude. As shown in Fig~\ref{fig:ellipticity_distribution}, the high-ellipticity subsample is preferentially composed of smaller halo voids, which might partly explain its weaker correlation with WL voids.
    \item Interestingly, rounder halo voids and whose major axes are less aligned with the LOS exhibit marginally stronger correlations. As shown in Fig.~\ref{fig:alignment_distribution}, we did not find any correlation between the angle of a halo void’s principal axis relative to the LOS and its angular size, $\theta_{hv}$. It is important to note, however, that the selection is also based on halo void ellipticity, which serves as a proxy for the elongation of their geometry. Consequently, the subsample consisting of more elongated halo voids with stronger LOS alignment is preferentially populated by smaller voids. Since smaller halo voids generally produce weaker lensing signals, this size dependence may contribute to the reduced correlation amplitude observed for that subsample.
\end{itemize}
To investigate whether the weaker correlations observed for the aligned and nonspherical halo void subsamples are driven primarily by their smaller characteristic radii, we repeated the analysis after dividing the halo void sample into two radius bins; namely, $R_{hv}< 40\ h^{-1}\,\mathrm{Mpc}$ and $R_{hv}\geq 40\ h^{-1}\,\mathrm{Mpc}$. The resulting cross-correlation profile is shown in Fig.~\ref{fig:split_tracer_voids_radius}. We find that the amplitudes of the central positive correlations are broadly similar for the two subsamples, suggesting that halo void size alone is unlikely to fully explain the morphology dependence discussed above. The largest differences appear near and beyond the compensation wall ($R/R_{wlv}\gtrsim 1.3$), where the smaller halo void subsample recovers more rapidly and becomes positively correlated, while the larger halo void subsample remains suppressed over a broader radial range. This suggests that the compensation wall region of WL voids is preferentially associated with smaller halo voids or subvoid-like structures, whereas large coherent halo voids are less common in these projected overdense environments.

Another factor that may contribute to the reduced correlation for the more elongated or LOS-aligned subsample is the increased likelihood of substructure within these voids. VIDE-defined 3D halo voids with high ellipticity often contain multiple internal density basins or significant deviations from a smooth radial profile. These internal complexities can disrupt the one-to-one mapping between 3D density depressions and the projected convergence field, further weakening the spatial overlap with WL voids. This behavior is consistent with the stacked density profiles shown in Fig. 7 of \citet{Schuster_2023}, where the least elliptical voids display the sharpest walls and deepest cores, while more elongated voids appear significantly smoother. This trend arises naturally from spherical averaging: when nearby regions of similar density are nonspherical, they intersect radial shells at different distances from the void center, effectively washing out the density contrasts. In contrast, rounder voids tend to have cleaner, more coherent interiors, which enhances their detectability in WL maps and strengthens their statistical association with WL-defined structures.

Taken together, these trends highlight that the WL-halo-void correspondence is not solely governed by void size or radial depth, but also by subtler aspects of void morphology and geometry. While the differences between the subsamples are modest, they point toward a consistent physical picture: WL voids preferentially align with large, relatively isotropic underdensities in the 3D matter field, whereas small or highly elongated structures contribute less efficiently to the WL signal. The latter occurs because elongated voids, especially those stretched along the LOS, produce projected density profiles that are shallower and more easily polluted by surrounding structures. Even when such voids correspond to true 3D minima, their lensing imprints are diluted due to the intrinsic projection effects of WL observables.

It is also worth noting that these morphological dependencies interact nontrivially with the redshift-dependent effects described earlier. Since the LOS projection length increases with redshift, any geometric complexity in 3D voids becomes increasingly entangled with intervening mass fluctuations, amplifying differences between spherical and elongated voids.

Finally, these subsample results emphasize that the common assumption adopted (i.e., treating WL voids as direct 2D projections of isotropic 3D voids) might be overly simplistic. Instead, the WL void population appears to select a biased subset of the full halo-void population, preferentially sampling voids with particular geometric and environmental properties. Understanding and modeling this selection function will be important for future cosmological applications that rely on WL void statistics, especially those targeting modified gravity or dark-energy signatures, where void shapes and orientations could play a nonnegligible role.

%%%%%%%%%%%%%%%%%%%%%%%%%%%%%%%%%%%%%%%%%%%%%%%%%%%%%%%%%%%%%%
\section{Conclusions}
\label{sect:conclusion}

In this work, we carried out a cross-correlation analysis between voids identified in halo tracer fields and those detected in WL convergence maps, using the T17 simulations and DES Y3 source redshift distributions as our testbed. Our measurements demonstrate that WL voids and 3D halo voids exhibit a clear and statistically significant spatial connection at small angular separations, particularly for halo voids at low redshifts ($0 < z_{hv} < 0.5$). This result establishes that WL voids, defined purely from the projected mass distribution, are effective tracers of true underdensities in the 3D matter field.

The amplitude of this correlation depends sensitively on how WL voids are selected. Catalogs constructed from unfiltered or positive-amplitude WL peaks yield robust, high-significance detections of WL–halo-void alignment, whereas imposing a high peak threshold ($\kappa > 0.01$) drastically weakens the cross-correlation. This arises because stringent amplitude cuts reduce the number of detectable peaks, producing fewer and larger WL voids whose projected geometry washes out finer correspondence with the 3D void population. This behavior highlights an important balance for observational analyses: increasing WL peak purity comes at the cost of degrading the connection to true matter underdensities.

Our measurements also reveal a strong redshift dependence in both void samples. For the DES Y3-like source redshift distributions adopted in this work, the correlation signal is largest for low-redshift halo voids ($z_{hv} \approx 0.3-0.5$); this is consistent with expectations from lensing theory, namely: the lensing kernel peaks at approximately half the distance to the source galaxies.
On the WL side, voids identified in the convergence maps constructed using higher-redshift tomographic bins exhibit stronger correlations with halo voids, reflecting the greater cumulative mass fluctuations projected onto these source planes.
The precise redshift range of maximum sensitivity, however, depends on the adopted source redshift distribution and WL map configuration. Therefore, it may vary across different surveys and tomographic selections. 

By further subdividing halo voids based on ellipticity and orientation, we find that morphological properties play a measurable, though modest, role in shaping the correlation. Rounder halo voids display slightly stronger correlations, in part because more elongated voids tend to be smaller and therefore contribute more weakly to the projected WL signal. Voids whose major axes are not as closely aligned with the LOS also correlate more strongly with WL voids. This is likely due to the expectation that LOS-elongated structures produce diluted projected underdensities unless they are sufficiently large or deep. These findings are consistent with the emerging picture that WL voids preferentially select projected underdensities associated with relatively large, round 3D voids; whereas voids with extreme shapes or alignments might require more careful modeling.

Our results illuminate the physical nature of WL voids and their relation to the underlying matter distribution. They also illustrate the importance of tracer selection, redshift weighting, and void morphology when interpreting WL void observables. A natural next step is to extend this analysis to real observational data such as from DES or forthcoming stage-IV surveys such as 
Rubin and Euclid, enabling a direct assessment of the WL-halo-void correspondence in observational conditions. Incorporating more sophisticated treatments of shape noise, photometric-redshift uncertainties, and selection effects will be essential for extracting cosmological information from WL voids. In particular, combining 3D halo void catalogs, WL voids, and density field reconstructions could help disentangle contributions from true underdensities and LOS projections, ultimately improving void-based constraints on gravity, dark energy, and the growth of structure.

%%%%%%%%%%%%%%%%%%%%%%%%%%%%%%%%%%%%%%%%%%%%%%%%%%%%%%%%%%%%%%
\begin{acknowledgements}
This research was supported by the DLR through its funding of the Euclid Science Ground Segment development team at LMU 
(DLR award 50QE2304) and by the Ludwig-Maximilians-Universit\"at in Munich.
\end{acknowledgements}

\bibliographystyle{aa}
\bibliography{example}

@ARTICLE{Healpix_ref,
       author = {{G{\'o}rski}, K.~M. and {Hivon}, E. and {Banday}, A.~J. and {Wandelt}, B.~D. and {Hansen}, F.~K. and {Reinecke}, M. and {Bartelmann}, M.},
        title = "{HEALPix: A Framework for High-Resolution Discretization and Fast Analysis of Data Distributed on the Sphere}",
      journal = {\apj},
         year = 2005,
        month = apr,
       volume = {622},
       number = {2},
        pages = {759-771},
          doi = {10.1086/427976},
archivePrefix = {arXiv},
       eprint = {astro-ph/0409513},
 primaryClass = {astro-ph},
       adsurl = {https://ui.adsabs.harvard.edu/abs/2005ApJ...622..759G}
}

@ARTICLE{Schuster_2023,
       author = {{Schuster}, Nico and {Hamaus}, Nico and {Dolag}, Klaus and {Weller}, Jochen},
        title = "{Why cosmic voids matter: nonlinear structure \& linear dynamics}",
      journal = {\jcap},
         year = 2023,
        month = may,
       volume = {2023},
       number = {5},
          eid = {031},
        pages = {031},
          doi = {10.1088/1475-7516/2023/05/031},
archivePrefix = {arXiv},
       eprint = {2210.02457},
 primaryClass = {astro-ph.CO},
       adsurl = {https://ui.adsabs.harvard.edu/abs/2023JCAP...05..031S}
}

@ARTICLE{Hamaus_2016,
       author = {{Hamaus}, Nico and {Pisani}, Alice and {Sutter}, P.~M. and {Lavaux}, Guilhem and {Escoffier}, St{\'e}phanie and {Wandelt}, Benjamin D. and {Weller}, Jochen},
        title = "{Constraints on Cosmology and Gravity from the Dynamics of Voids}",
      journal = {\prl},
         year = 2016,
        month = aug,
       volume = {117},
       number = {9},
          eid = {091302},
        pages = {091302},
          doi = {10.1103/PhysRevLett.117.091302},
archivePrefix = {arXiv},
       eprint = {1602.01784},
 primaryClass = {astro-ph.CO},
       adsurl = {https://ui.adsabs.harvard.edu/abs/2016PhRvL.117i1302H}
}

@ARTICLE{Hartlap_2007,
       author = {{Hartlap}, J. and {Simon}, P. and {Schneider}, P.},
        title = "{Why your model parameter confidences might be too optimistic. Unbiased estimation of the inverse covariance matrix}",
      journal = {\aap},
         year = 2007,
        month = mar,
       volume = {464},
       number = {1},
        pages = {399-404},
          doi = {10.1051/0004-6361:20066170},
archivePrefix = {arXiv},
       eprint = {astro-ph/0608064},
 primaryClass = {astro-ph},
       adsurl = {https://ui.adsabs.harvard.edu/abs/2007A&A...464..399H}
}

@ARTICLE{Ferlito_2025,
       author = {{Ferlito}, Fulvio and {Springel}, Volker and {Davies}, Christopher T. and {Kurita}, Toshiki and {Delgado}, Ana Maria and {Bose}, Sownak and {Hernquist}, Lars},
        title = "{Fully non-linear simulations of galaxy intrinsic alignments for weak lensing with the MillenniumTNG lightcone}",
      journal = {\mnras},
         year = 2025,
        month = oct,
          doi = {10.1093/mnras/staf1722},
archivePrefix = {arXiv},
       eprint = {2505.15882},
 primaryClass = {astro-ph.CO},
       adsurl = {https://ui.adsabs.harvard.edu/abs/2025MNRAS.tmp.1619F}
}

@ARTICLE{Davies_2019b,
       author = {{Davies}, Christopher T. and {Cautun}, Marius and {Li}, Baojiu},
        title = "{Cosmological test of gravity using weak lensing voids}",
      journal = {\mnras},
         year = 2019,
        month = dec,
       volume = {490},
       number = {4},
        pages = {4907-4917},
          doi = {10.1093/mnras/stz2933},
archivePrefix = {arXiv},
       eprint = {1907.06657},
 primaryClass = {astro-ph.CO},
       adsurl = {https://ui.adsabs.harvard.edu/abs/2019MNRAS.490.4907D}
}

@ARTICLE{Davies_2018,
       author = {{Davies}, Christopher T. and {Cautun}, Marius and {Li}, Baojiu},
        title = "{Weak lensing by voids in weak lensing maps}",
      journal = {\mnras},
         year = 2018,
        month = oct,
       volume = {480},
       number = {1},
        pages = {L101-L105},
          doi = {10.1093/mnrasl/sly135},
archivePrefix = {arXiv},
       eprint = {1803.08717},
 primaryClass = {astro-ph.CO},
       adsurl = {https://ui.adsabs.harvard.edu/abs/2018MNRAS.480L.101D}
}

@ARTICLE{Oguri_2018_a,
       author = {{Oguri}, Masamune and {Lin}, Yen-Ting and {Lin}, Sheng-Chieh and {Nishizawa}, Atsushi J. and {More}, Anupreeta and {More}, Surhud and {Hsieh}, Bau-Ching and {Medezinski}, Elinor and {Miyatake}, Hironao and {Jian}, Hung-Yu and {Lin}, Lihwai and {Takada}, Masahiro and {Okabe}, Nobuhiro and {Speagle}, Joshua S. and {Coupon}, Jean and {Leauthaud}, Alexie and {Lupton}, Robert H. and {Miyazaki}, Satoshi and {Price}, Paul A. and {Tanaka}, Masayuki and {Chiu}, I. -Non and {Komiyama}, Yutaka and {Okura}, Yuki and {Tanaka}, Manobu M. and {Usuda}, Tomonori},
        title = "{An optically-selected cluster catalog at redshift 0.1 < z < 1.1 from the Hyper Suprime-Cam Subaru Strategic Program S16A data}",
      journal = {\pasj},
         year = 2018,
        month = jan,
       volume = {70},
          eid = {S20},
        pages = {S20},
          doi = {10.1093/pasj/psx042},
archivePrefix = {arXiv},
       eprint = {1701.00818},
 primaryClass = {astro-ph.CO},
       adsurl = {https://ui.adsabs.harvard.edu/abs/2018PASJ...70S..20O}
}

@ARTICLE{Reid_2016,
       author = {{Reid}, Beth and {Ho}, Shirley and {Padmanabhan}, Nikhil and {Percival}, Will J. and {Tinker}, Jeremy and {Tojeiro}, Rita and {White}, Martin and {Eisenstein}, Daniel J. and {Maraston}, Claudia and {Ross}, Ashley J. and {S{\'a}nchez}, Ariel G. and {Schlegel}, David and {Sheldon}, Erin and {Strauss}, Michael A. and {Thomas}, Daniel and {Wake}, David and {Beutler}, Florian and {Bizyaev}, Dmitry and {Bolton}, Adam S. and {Brownstein}, Joel R. and {Chuang}, Chia-Hsun and {Dawson}, Kyle and {Harding}, Paul and {Kitaura}, Francisco-Shu and {Leauthaud}, Alexie and {Masters}, Karen and {McBride}, Cameron K. and {More}, Surhud and {Olmstead}, Matthew D. and {Oravetz}, Daniel and {Nuza}, Sebasti{\'a}n E. and {Pan}, Kaike and {Parejko}, John and {Pforr}, Janine and {Prada}, Francisco and {Rodr{\'\i}guez-Torres}, Sergio and {Salazar-Albornoz}, Salvador and {Samushia}, Lado and {Schneider}, Donald P. and {Sc{\'o}ccola}, Claudia G. and {Simmons}, Audrey and {Vargas-Magana}, Mariana},
        title = "{SDSS-III Baryon Oscillation Spectroscopic Survey Data Release 12: galaxy target selection and large-scale structure catalogues}",
      journal = {\mnras},
         year = 2016,
        month = jan,
       volume = {455},
       number = {2},
        pages = {1553-1573},
          doi = {10.1093/mnras/stv2382},
archivePrefix = {arXiv},
       eprint = {1509.06529},
 primaryClass = {astro-ph.CO},
       adsurl = {https://ui.adsabs.harvard.edu/abs/2016MNRAS.455.1553R}
}

@ARTICLE{Oguri_2018,
       author = {{Oguri}, Masamune and {Miyazaki}, Satoshi and {Hikage}, Chiaki and {Mandelbaum}, Rachel and {Utsumi}, Yousuke and {Miyatake}, Hironao and {Takada}, Masahiro and {Armstrong}, Robert and {Bosch}, James and {Komiyama}, Yutaka and {Leauthaud}, Alexie and {More}, Surhud and {Nishizawa}, Atsushi J. and {Okabe}, Nobuhiro and {Tanaka}, Masayuki},
        title = "{Two- and three-dimensional wide-field weak lensing mass maps from the Hyper Suprime-Cam Subaru Strategic Program S16A data}",
      journal = {\pasj},
         year = 2018,
        month = jan,
       volume = {70},
          eid = {S26},
        pages = {S26},
          doi = {10.1093/pasj/psx070},
archivePrefix = {arXiv},
       eprint = {1705.06792},
 primaryClass = {astro-ph.CO},
       adsurl = {https://ui.adsabs.harvard.edu/abs/2018PASJ...70S..26O}
}

@ARTICLE{Sheth_2004,
       author = {{Sheth}, Ravi K. and {van de Weygaert}, Rien},
        title = "{A hierarchy of voids: much ado about nothing}",
      journal = {\mnras},
         year = 2004,
        month = may,
       volume = {350},
       number = {2},
        pages = {517-538},
          doi = {10.1111/j.1365-2966.2004.07661.x},
archivePrefix = {arXiv},
       eprint = {astro-ph/0311260},
 primaryClass = {astro-ph},
       adsurl = {https://ui.adsabs.harvard.edu/abs/2004MNRAS.350..517S}
}

@ARTICLE{Pisani_2019,
       author = {{Pisani}, Alice and {Massara}, Elena and {Spergel}, David N. and {Alonso}, David and {Baker}, Tessa and {Cai}, Yan-Chuan and {Cautun}, Marius and {Davies}, Christopher and {Demchenko}, Vasiliy and {Dor{\'e}}, Olivier and {Goulding}, Andy and {Habouzit}, M{\'e}lanie and {Hamaus}, Nico and {Hawken}, Adam and {Hirata}, Christopher M. and {Ho}, Shirley and {Jain}, Bhuvnesh and {Kreisch}, Christina D. and {Marulli}, Federico and {Padilla}, Nelson and {Pollina}, Giorgia and {Sahl{\'e}n}, Martin and {Sheth}, Ravi K. and {Somerville}, Rachel and {Szapudi}, Istvan and {van de Weygaert}, Rien and {Villaescusa-Navarro}, Francisco and {Wandelt}, Benjamin D. and {Wang}, Yun},
        title = "{Cosmic voids: a novel probe to shed light on our Universe}",
      journal = {\baas},
         year = 2019,
        month = may,
       volume = {51},
       number = {3},
          eid = {40},
        pages = {40},
          doi = {10.48550/arXiv.1903.05161},
archivePrefix = {arXiv},
       eprint = {1903.05161},
 primaryClass = {astro-ph.CO},
       adsurl = {https://ui.adsabs.harvard.edu/abs/2019BAAS...51c..40P}
}

@ARTICLE{Cautun_2018,
       author = {{Cautun}, Marius and {Paillas}, Enrique and {Cai}, Yan-Chuan and {Bose}, Sownak and {Armijo}, Joaquin and {Li}, Baojiu and {Padilla}, Nelson},
        title = "{The Santiago-Harvard-Edinburgh-Durham void comparison - I. SHEDding light on chameleon gravity tests}",
      journal = {\mnras},
         year = 2018,
        month = may,
       volume = {476},
       number = {3},
        pages = {3195-3217},
          doi = {10.1093/mnras/sty463},
archivePrefix = {arXiv},
       eprint = {1710.01730},
 primaryClass = {astro-ph.CO},
       adsurl = {https://ui.adsabs.harvard.edu/abs/2018MNRAS.476.3195C}
}

@ARTICLE{Hamaus_2020,
       author = {{Hamaus}, Nico and {Pisani}, Alice and {Choi}, Jin-Ah and {Lavaux}, Guilhem and {Wandelt}, Benjamin D. and {Weller}, Jochen},
        title = "{Precision cosmology with voids in the final BOSS data}",
      journal = {\jcap},
         year = 2020,
        month = dec,
       volume = {2020},
       number = {12},
          eid = {023},
        pages = {023},
          doi = {10.1088/1475-7516/2020/12/023},
archivePrefix = {arXiv},
       eprint = {2007.07895},
 primaryClass = {astro-ph.CO},
       adsurl = {https://ui.adsabs.harvard.edu/abs/2020JCAP...12..023H}
}

@ARTICLE{Nadathur_2019,
       author = {{Nadathur}, Seshadri and {Percival}, Will J.},
        title = "{An accurate linear model for redshift space distortions in the void-galaxy correlation function}",
      journal = {\mnras},
         year = 2019,
        month = mar,
       volume = {483},
       number = {3},
        pages = {3472-3487},
          doi = {10.1093/mnras/sty3372},
archivePrefix = {arXiv},
       eprint = {1712.07575},
 primaryClass = {astro-ph.CO},
       adsurl = {https://ui.adsabs.harvard.edu/abs/2019MNRAS.483.3472N}
}

@ARTICLE{Correa_2021,
       author = {{Correa}, Carlos M. and {Paz}, Dante J. and {S{\'a}nchez}, Ariel G. and {Ruiz}, Andr{\'e}s N. and {Padilla}, Nelson D. and {Angulo}, Ra{\'u}l E.},
        title = "{Redshift-space effects in voids and their impact on cosmological tests. Part I: the void size function}",
      journal = {\mnras},
         year = 2021,
        month = jan,
       volume = {500},
       number = {1},
        pages = {911-925},
          doi = {10.1093/mnras/staa3252},
archivePrefix = {arXiv},
       eprint = {2007.12064},
 primaryClass = {astro-ph.CO},
       adsurl = {https://ui.adsabs.harvard.edu/abs/2021MNRAS.500..911C}
}

@ARTICLE{Fang_2019,
       author = {{Fang}, Y. and {Hamaus}, N. and {Jain}, B. and {Pandey}, S. and {Pollina}, G. and {S{\'a}nchez}, C. and {Kov{\'a}cs}, A. and {Chang}, C. and {Carretero}, J. and {Castander}, F.~J. and {Choi}, A. and {Crocce}, M. and {DeRose}, J. and {Fosalba}, P. and {Gatti}, M. and {Gazta{\~n}aga}, E. and {Gruen}, D. and {Hartley}, W.~G. and {Hoyle}, B. and {MacCrann}, N. and {Prat}, J. and {Rau}, M.~M. and {Rykoff}, E.~S. and {Samuroff}, S. and {Sheldon}, E. and {Troxel}, M.~A. and {Vielzeuf}, P. and {Zuntz}, J. and {Annis}, J. and {Avila}, S. and {Bertin}, E. and {Brooks}, D. and {Burke}, D.~L. and {Carnero Rosell}, A. and {Carrasco Kind}, M. and {Cawthon}, R. and {da Costa}, L.~N. and {De Vicente}, J. and {Desai}, S. and {Diehl}, H.~T. and {Dietrich}, J.~P. and {Doel}, P. and {Everett}, S. and {Evrard}, A.~E. and {Flaugher}, B. and {Frieman}, J. and {Garc{\'\i}a-Bellido}, J. and {Gerdes}, D.~W. and {Gruendl}, R.~A. and {Gutierrez}, G. and {Hollowood}, D.~L. and {James}, D.~J. and {Jarvis}, M. and {Kuropatkin}, N. and {Lahav}, O. and {Maia}, M.~A.~G. and {Marshall}, J.~L. and {Melchior}, P. and {Menanteau}, F. and {Miquel}, R. and {Palmese}, A. and {Plazas}, A.~A. and {Romer}, A.~K. and {Roodman}, A. and {Sanchez}, E. and {Serrano}, S. and {Sevilla-Noarbe}, I. and {Smith}, M. and {Soares-Santos}, M. and {Sobreira}, F. and {Suchyta}, E. and {Swanson}, M.~E.~C. and {Tarle}, G. and {Thomas}, D. and {Vikram}, V. and {Walker}, A.~R. and {Weller}, J. and {DES Collaboration}},
        title = "{Dark Energy Survey year 1 results: the relationship between mass and light around cosmic voids}",
      journal = {\mnras},
         year = 2019,
        month = dec,
       volume = {490},
       number = {3},
        pages = {3573-3587},
          doi = {10.1093/mnras/stz2805},
archivePrefix = {arXiv},
       eprint = {1909.01386},
 primaryClass = {astro-ph.CO},
       adsurl = {https://ui.adsabs.harvard.edu/abs/2019MNRAS.490.3573F}
}

@ARTICLE{Pollina_2017,
       author = {{Pollina}, Giorgia and {Hamaus}, Nico and {Dolag}, Klaus and {Weller}, Jochen and {Baldi}, Marco and {Moscardini}, Lauro},
        title = "{On the linearity of tracer bias around voids}",
      journal = {\mnras},
         year = 2017,
        month = jul,
       volume = {469},
       number = {1},
        pages = {787-799},
          doi = {10.1093/mnras/stx785},
archivePrefix = {arXiv},
       eprint = {1610.06176},
 primaryClass = {astro-ph.CO},
       adsurl = {https://ui.adsabs.harvard.edu/abs/2017MNRAS.469..787P}
}

@ARTICLE{Cai_2016,
       author = {{Cai}, Yan-Chuan and {Taylor}, Andy and {Peacock}, John A. and {Padilla}, Nelson},
        title = "{Redshift-space distortions around voids}",
      journal = {\mnras},
         year = 2016,
        month = nov,
       volume = {462},
       number = {3},
        pages = {2465-2477},
          doi = {10.1093/mnras/stw1809},
archivePrefix = {arXiv},
       eprint = {1603.05184},
 primaryClass = {astro-ph.CO},
       adsurl = {https://ui.adsabs.harvard.edu/abs/2016MNRAS.462.2465C}
}

@ARTICLE{Gruen_2016,
       author = {{Gruen}, D. and {Friedrich}, O. and {Amara}, A. and {Bacon}, D. and {Bonnett}, C. and {Hartley}, W. and {Jain}, B. and {Jarvis}, M. and {Kacprzak}, T. and {Krause}, E. and {Mana}, A. and {Rozo}, E. and {Rykoff}, E.~S. and {Seitz}, S. and {Sheldon}, E. and {Troxel}, M.~A. and {Vikram}, V. and {Abbott}, T.~M.~C. and {Abdalla}, F.~B. and {Allam}, S. and {Armstrong}, R. and {Banerji}, M. and {Bauer}, A.~H. and {Becker}, M.~R. and {Benoit-L{\'e}vy}, A. and {Bernstein}, G.~M. and {Bernstein}, R.~A. and {Bertin}, E. and {Bridle}, S.~L. and {Brooks}, D. and {Buckley-Geer}, E. and {Burke}, D.~L. and {Capozzi}, D. and {Carnero Rosell}, A. and {Carrasco Kind}, M. and {Carretero}, J. and {Crocce}, M. and {Cunha}, C.~E. and {D'Andrea}, C.~B. and {da Costa}, L.~N. and {DePoy}, D.~L. and {Desai}, S. and {Diehl}, H.~T. and {Dietrich}, J.~P. and {Doel}, P. and {Eifler}, T.~F. and {Neto}, A. Fausti and {Fernandez}, E. and {Flaugher}, B. and {Fosalba}, P. and {Frieman}, J. and {Gerdes}, D.~W. and {Gruendl}, R.~A. and {Gutierrez}, G. and {Honscheid}, K. and {James}, D.~J. and {Kuehn}, K. and {Kuropatkin}, N. and {Lahav}, O. and {Li}, T.~S. and {Lima}, M. and {Maia}, M.~A.~G. and {March}, M. and {Martini}, P. and {Melchior}, P. and {Miller}, C.~J. and {Miquel}, R. and {Mohr}, J.~J. and {Nord}, B. and {Ogando}, R. and {Plazas}, A.~A. and {Reil}, K. and {Romer}, A.~K. and {Roodman}, A. and {Sako}, M. and {Sanchez}, E. and {Scarpine}, V. and {Schubnell}, M. and {Sevilla-Noarbe}, I. and {Smith}, R.~C. and {Soares-Santos}, M. and {Sobreira}, F. and {Suchyta}, E. and {Swanson}, M.~E.~C. and {Tarle}, G. and {Thaler}, J. and {Thomas}, D. and {Walker}, A.~R. and {Wechsler}, R.~H. and {Weller}, J. and {Zhang}, Y. and {Zuntz}, J.},
        title = "{Weak lensing by galaxy troughs in DES Science Verification data}",
      journal = {\mnras},
         year = 2016,
        month = jan,
       volume = {455},
       number = {3},
        pages = {3367-3380},
          doi = {10.1093/mnras/stv2506},
archivePrefix = {arXiv},
       eprint = {1507.05090},
 primaryClass = {astro-ph.CO},
       adsurl = {https://ui.adsabs.harvard.edu/abs/2016MNRAS.455.3367G}
}

@ARTICLE{Clampitt_2015,
       author = {{Clampitt}, Joseph and {Jain}, Bhuvnesh},
        title = "{Lensing measurements of the mass distribution in SDSS voids}",
      journal = {\mnras},
         year = 2015,
        month = dec,
       volume = {454},
       number = {4},
        pages = {3357-3365},
          doi = {10.1093/mnras/stv2215},
archivePrefix = {arXiv},
       eprint = {1404.1834},
 primaryClass = {astro-ph.CO},
       adsurl = {https://ui.adsabs.harvard.edu/abs/2015MNRAS.454.3357C}
}

@ARTICLE{Paz_2023,
       author = {{Paz}, Dante J. and {Correa}, Carlos M. and {Gualpa}, Sebasti{\'\i}n R. and {Ruiz}, Andres N. and {Bederi{\'a}n}, Carlos S. and {Gra{\~n}a}, R. Dario and {Padilla}, Nelson D.},
        title = "{Guess the cheese flavour by the size of its holes: a cosmological test using the abundance of popcorn voids}",
      journal = {\mnras},
         year = 2023,
        month = jun,
       volume = {522},
       number = {2},
        pages = {2553-2569},
          doi = {10.1093/mnras/stad1146},
archivePrefix = {arXiv},
       eprint = {2212.06849},
 primaryClass = {astro-ph.CO},
       adsurl = {https://ui.adsabs.harvard.edu/abs/2023MNRAS.522.2553P}
}

@ARTICLE{Colberg_2008,
       author = {{Colberg}, J{\"o}rg M. and {Pearce}, Frazer and {Foster}, Caroline and {Platen}, Erwin and {Brunino}, Riccardo and {Neyrinck}, Mark and {Basilakos}, Spyros and {Fairall}, Anthony and {Feldman}, Hume and {Gottl{\"o}ber}, Stefan and {Hahn}, Oliver and {Hoyle}, Fiona and {M{\"u}ller}, Volker and {Nelson}, Lorne and {Plionis}, Manolis and {Porciani}, Cristiano and {Shandarin}, Sergei and {Vogeley}, Michael S. and {van de Weygaert}, Rien},
        title = "{The Aspen-Amsterdam void finder comparison project}",
      journal = {\mnras},
         year = 2008,
        month = jun,
       volume = {387},
       number = {2},
        pages = {933-944},
          doi = {10.1111/j.1365-2966.2008.13307.x},
archivePrefix = {arXiv},
       eprint = {0803.0918},
 primaryClass = {astro-ph},
       adsurl = {https://ui.adsabs.harvard.edu/abs/2008MNRAS.387..933C}
}

@ARTICLE{Platen_2007,
       author = {{Platen}, Erwin and {van de Weygaert}, Rien and {Jones}, Bernard J.~T.},
        title = "{A cosmic watershed: the WVF void detection technique}",
      journal = {\mnras},
         year = 2007,
        month = sep,
       volume = {380},
       number = {2},
        pages = {551-570},
          doi = {10.1111/j.1365-2966.2007.12125.x},
archivePrefix = {arXiv},
       eprint = {0706.2788},
 primaryClass = {astro-ph},
       adsurl = {https://ui.adsabs.harvard.edu/abs/2007MNRAS.380..551P}
}

@ARTICLE{Kaiser_1984,
       author = {{Kaiser}, N.},
        title = "{On the spatial correlations of Abell clusters.}",
      journal = {\apjl},
         year = 1984,
        month = sep,
       volume = {284},
        pages = {L9-L12},
          doi = {10.1086/184341},
       adsurl = {https://ui.adsabs.harvard.edu/abs/1984ApJ...284L...9K}
}

@BOOK{Peebles_1980,
       author = {{Peebles}, P.~J.~E.},
        title = "{The large-scale structure of the universe}",
         year = 1980,
       adsurl = {https://ui.adsabs.harvard.edu/abs/1980lssu.book.....P}
}

@BOOK{Mo_2010,
       author = {{Mo}, Houjun and {van den Bosch}, Frank C. and {White}, Simon},
        title = "{Galaxy Formation and Evolution}",
         year = 2010,
          doi = {10.1017/CBO9780511807244},
       adsurl = {https://ui.adsabs.harvard.edu/abs/2010gfe..book.....M}
}

@ARTICLE{White_1978,
       author = {{White}, S.~D.~M. and {Rees}, M.~J.},
        title = "{Core condensation in heavy halos: a two-stage theory for galaxy formation and clustering.}",
      journal = {\mnras},
         year = 1978,
        month = may,
       volume = {183},
        pages = {341-358},
          doi = {10.1093/mnras/183.3.341},
       adsurl = {https://ui.adsabs.harvard.edu/abs/1978MNRAS.183..341W}
}

@ARTICLE{DES_2016,
       author = {{Dark Energy Survey Collaboration} and {Abbott}, T. and {Abdalla}, F.~B. and {Aleksi{\'c}}, J. and {Allam}, S. and {Amara}, A. and {Bacon}, D. and {Balbinot}, E. and {Banerji}, M. and {Bechtol}, K. and {Benoit-L{\'e}vy}, A. and {Bernstein}, G.~M. and {Bertin}, E. and {Blazek}, J. and {Bonnett}, C. and {Bridle}, S. and {Brooks}, D. and {Brunner}, R.~J. and {Buckley-Geer}, E. and {Burke}, D.~L. and {Caminha}, G.~B. and {Capozzi}, D. and {Carlsen}, J. and {Carnero-Rosell}, A. and {Carollo}, M. and {Carrasco-Kind}, M. and {Carretero}, J. and {Castander}, F.~J. and {Clerkin}, L. and {Collett}, T. and {Conselice}, C. and {Crocce}, M. and {Cunha}, C.~E. and {D'Andrea}, C.~B. and {da Costa}, L.~N. and {Davis}, T.~M. and {Desai}, S. and {Diehl}, H.~T. and {Dietrich}, J.~P. and {Dodelson}, S. and {Doel}, P. and {Drlica-Wagner}, A. and {Estrada}, J. and {Etherington}, J. and {Evrard}, A.~E. and {Fabbri}, J. and {Finley}, D.~A. and {Flaugher}, B. and {Foley}, R.~J. and {Fosalba}, P. and {Frieman}, J. and {Garc{\'\i}a-Bellido}, J. and {Gaztanaga}, E. and {Gerdes}, D.~W. and {Giannantonio}, T. and {Goldstein}, D.~A. and {Gruen}, D. and {Gruendl}, R.~A. and {Guarnieri}, P. and {Gutierrez}, G. and {Hartley}, W. and {Honscheid}, K. and {Jain}, B. and {James}, D.~J. and {Jeltema}, T. and {Jouvel}, S. and {Kessler}, R. and {King}, A. and {Kirk}, D. and {Kron}, R. and {Kuehn}, K. and {Kuropatkin}, N. and {Lahav}, O. and {Li}, T.~S. and {Lima}, M. and {Lin}, H. and {Maia}, M.~A.~G. and {Makler}, M. and {Manera}, M. and {Maraston}, C. and {Marshall}, J.~L. and {Martini}, P. and {McMahon}, R.~G. and {Melchior}, P. and {Merson}, A. and {Miller}, C.~J. and {Miquel}, R. and {Mohr}, J.~J. and {Morice-Atkinson}, X. and {Naidoo}, K. and {Neilsen}, E. and {Nichol}, R.~C. and {Nord}, B. and {Ogando}, R. and {Ostrovski}, F. and {Palmese}, A. and {Papadopoulos}, A. and {Peiris}, H.~V. and {Peoples}, J. and {Percival}, W.~J. and {Plazas}, A.~A. and {Reed}, S.~L. and {Refregier}, A. and {Romer}, A.~K. and {Roodman}, A. and {Ross}, A. and {Rozo}, E. and {Rykoff}, E.~S. and {Sadeh}, I. and {Sako}, M. and {S{\'a}nchez}, C. and {Sanchez}, E. and {Santiago}, B. and {Scarpine}, V. and {Schubnell}, M. and {Sevilla-Noarbe}, I. and {Sheldon}, E. and {Smith}, M. and {Smith}, R.~C. and {Soares-Santos}, M. and {Sobreira}, F. and {Soumagnac}, M. and {Suchyta}, E. and {Sullivan}, M. and {Swanson}, M. and {Tarle}, G. and {Thaler}, J. and {Thomas}, D. and {Thomas}, R.~C. and {Tucker}, D. and {Vieira}, J.~D. and {Vikram}, V. and {Walker}, A.~R. and {Wechsler}, R.~H. and {Weller}, J. and {Wester}, W. and {Whiteway}, L. and {Wilcox}, H. and {Yanny}, B. and {Zhang}, Y. and {Zuntz}, J.},
        title = "{The Dark Energy Survey: more than dark energy - an overview}",
      journal = {\mnras},
         year = 2016,
        month = aug,
       volume = {460},
       number = {2},
        pages = {1270-1299},
          doi = {10.1093/mnras/stw641},
archivePrefix = {arXiv},
       eprint = {1601.00329},
 primaryClass = {astro-ph.CO},
       adsurl = {https://ui.adsabs.harvard.edu/abs/2016MNRAS.460.1270D}
}

@ARTICLE{Mao_2017,
       author = {{Mao}, Qingqing and {Berlind}, Andreas A. and {Scherrer}, Robert J. and {Neyrinck}, Mark C. and {Scoccimarro}, Rom{\'a}n and {Tinker}, Jeremy L. and {McBride}, Cameron K. and {Schneider}, Donald P. and {Pan}, Kaike and {Bizyaev}, Dmitry and {Malanushenko}, Elena and {Malanushenko}, Viktor},
        title = "{A Cosmic Void Catalog of SDSS DR12 BOSS Galaxies}",
      journal = {\apj},
         year = 2017,
        month = feb,
       volume = {835},
       number = {2},
          eid = {161},
        pages = {161},
          doi = {10.3847/1538-4357/835/2/161},
archivePrefix = {arXiv},
       eprint = {1602.02771},
 primaryClass = {astro-ph.CO},
       adsurl = {https://ui.adsabs.harvard.edu/abs/2017ApJ...835..161M}
}

@ARTICLE{Sanchez_2017,
       author = {{S{\'a}nchez}, C. and {Clampitt}, J. and {Kovacs}, A. and {Jain}, B. and {Garc{\'\i}a-Bellido}, J. and {Nadathur}, S. and {Gruen}, D. and {Hamaus}, N. and {Huterer}, D. and {Vielzeuf}, P. and {Amara}, A. and {Bonnett}, C. and {DeRose}, J. and {Hartley}, W.~G. and {Jarvis}, M. and {Lahav}, O. and {Miquel}, R. and {Rozo}, E. and {Rykoff}, E.~S. and {Sheldon}, E. and {Wechsler}, R.~H. and {Zuntz}, J. and {Abbott}, T.~M.~C. and {Abdalla}, F.~B. and {Annis}, J. and {Benoit-L{\'e}vy}, A. and {Bernstein}, G.~M. and {Bernstein}, R.~A. and {Bertin}, E. and {Brooks}, D. and {Buckley-Geer}, E. and {Carnero Rosell}, A. and {Carrasco Kind}, M. and {Carretero}, J. and {Crocce}, M. and {Cunha}, C.~E. and {D'Andrea}, C.~B. and {da Costa}, L.~N. and {Desai}, S. and {Diehl}, H.~T. and {Dietrich}, J.~P. and {Doel}, P. and {Evrard}, A.~E. and {Fausti Neto}, A. and {Flaugher}, B. and {Fosalba}, P. and {Frieman}, J. and {Gaztanaga}, E. and {Gruendl}, R.~A. and {Gutierrez}, G. and {Honscheid}, K. and {James}, D.~J. and {Krause}, E. and {Kuehn}, K. and {Lima}, M. and {Maia}, M.~A.~G. and {Marshall}, J.~L. and {Melchior}, P. and {Plazas}, A.~A. and {Reil}, K. and {Romer}, A.~K. and {Sanchez}, E. and {Schubnell}, M. and {Sevilla-Noarbe}, I. and {Smith}, R.~C. and {Soares-Santos}, M. and {Sobreira}, F. and {Suchyta}, E. and {Tarle}, G. and {Thomas}, D. and {Walker}, A.~R. and {Weller}, J. and {DES Collaboration}},
        title = "{Cosmic voids and void lensing in the Dark Energy Survey Science Verification data}",
      journal = {\mnras},
         year = 2017,
        month = feb,
       volume = {465},
       number = {1},
        pages = {746-759},
          doi = {10.1093/mnras/stw2745},
archivePrefix = {arXiv},
       eprint = {1605.03982},
 primaryClass = {astro-ph.CO},
       adsurl = {https://ui.adsabs.harvard.edu/abs/2017MNRAS.465..746S}
}

@ARTICLE{Sutter_2012,
       author = {{Sutter}, P.~M. and {Lavaux}, Guilhem and {Wandelt}, Benjamin D. and {Weinberg}, David H.},
        title = "{A Public Void Catalog from the SDSS DR7 Galaxy Redshift Surveys Based on the Watershed Transform}",
      journal = {\apj},
         year = 2012,
        month = dec,
       volume = {761},
       number = {1},
          eid = {44},
        pages = {44},
          doi = {10.1088/0004-637X/761/1/44},
archivePrefix = {arXiv},
       eprint = {1207.2524},
 primaryClass = {astro-ph.CO},
       adsurl = {https://ui.adsabs.harvard.edu/abs/2012ApJ...761...44S}
}

@ARTICLE{Hoyle_2012,
       author = {{Hoyle}, Fiona and {Vogeley}, M.~S. and {Pan}, D.},
        title = "{Photometric properties of void galaxies in the Sloan Digital Sky Survey Data Release 7}",
      journal = {\mnras},
         year = 2012,
        month = nov,
       volume = {426},
       number = {4},
        pages = {3041-3050},
          doi = {10.1111/j.1365-2966.2012.21943.x},
archivePrefix = {arXiv},
       eprint = {1205.1843},
 primaryClass = {astro-ph.CO},
       adsurl = {https://ui.adsabs.harvard.edu/abs/2012MNRAS.426.3041H}
}

@ARTICLE{Pan_2012,
       author = {{Pan}, Danny C. and {Vogeley}, Michael S. and {Hoyle}, Fiona and {Choi}, Yun-Young and {Park}, Changbom},
        title = "{Cosmic voids in Sloan Digital Sky Survey Data Release 7}",
      journal = {\mnras},
         year = 2012,
        month = apr,
       volume = {421},
       number = {2},
        pages = {926-934},
          doi = {10.1111/j.1365-2966.2011.20197.x},
archivePrefix = {arXiv},
       eprint = {1103.4156},
 primaryClass = {astro-ph.CO},
       adsurl = {https://ui.adsabs.harvard.edu/abs/2012MNRAS.421..926P}
}

@ARTICLE{Davies_2021,
       author = {{Davies}, Christopher T. and {Cautun}, Marius and {Giblin}, Benjamin and {Li}, Baojiu and {Harnois-D{\'e}raps}, Joachim and {Cai}, Yan-Chuan},
        title = "{Constraining cosmology with weak lensing voids}",
      journal = {\mnras},
         year = 2021,
        month = oct,
       volume = {507},
       number = {2},
        pages = {2267-2282},
          doi = {10.1093/mnras/stab2251},
archivePrefix = {arXiv},
       eprint = {2010.11954},
 primaryClass = {astro-ph.CO},
       adsurl = {https://ui.adsabs.harvard.edu/abs/2021MNRAS.507.2267D}
}

@ARTICLE{Hamaus_2014,
       author = {{Hamaus}, Nico and {Sutter}, P.~M. and {Wandelt}, Benjamin D.},
        title = "{Universal Density Profile for Cosmic Voids}",
      journal = {\prl},
         year = 2014,
        month = jun,
       volume = {112},
       number = {25},
          eid = {251302},
        pages = {251302},
          doi = {10.1103/PhysRevLett.112.251302},
archivePrefix = {arXiv},
       eprint = {1403.5499},
 primaryClass = {astro-ph.CO},
       adsurl = {https://ui.adsabs.harvard.edu/abs/2014PhRvL.112y1302H}
}

@ARTICLE{Shimasue_2024,
       author = {{Shimasue}, Takumi and {Osato}, Ken and {Oguri}, Masamune and {Shimakawa}, Rhythm and {Nishizawa}, Atsushi J.},
        title = "{Line-of-sight structure of troughs identified in Subaru Hyper Suprime-Cam Year 3 weak lensing mass maps}",
      journal = {\mnras},
         year = 2024,
        month = jan,
       volume = {527},
       number = {3},
        pages = {5974-5987},
          doi = {10.1093/mnras/stad3542},
archivePrefix = {arXiv},
       eprint = {2307.11407},
 primaryClass = {astro-ph.CO},
       adsurl = {https://ui.adsabs.harvard.edu/abs/2024MNRAS.527.5974S}
}

@ARTICLE{Gatti_2021,
       author = {{Gatti}, M. and {Sheldon}, E. and {Amon}, A. and {Becker}, M. and {Troxel}, M. and {Choi}, A. and {Doux}, C. and {MacCrann}, N. and {Navarro-Alsina}, A. and {Harrison}, I. and {Gruen}, D. and {Bernstein}, G. and {Jarvis}, M. and {Secco}, L.~F. and {Fert{\'e}}, A. and {Shin}, T. and {McCullough}, J. and {Rollins}, R.~P. and {Chen}, R. and {Chang}, C. and {Pandey}, S. and {Tutusaus}, I. and {Prat}, J. and {Elvin-Poole}, J. and {Sanchez}, C. and {Plazas}, A.~A. and {Roodman}, A. and {Zuntz}, J. and {Abbott}, T.~M.~C. and {Aguena}, M. and {Allam}, S. and {Annis}, J. and {Avila}, S. and {Bacon}, D. and {Bertin}, E. and {Bhargava}, S. and {Brooks}, D. and {Burke}, D.~L. and {Carnero Rosell}, A. and {Carrasco Kind}, M. and {Carretero}, J. and {Castander}, F.~J. and {Conselice}, C. and {Costanzi}, M. and {Crocce}, M. and {da Costa}, L.~N. and {Davis}, T.~M. and {De Vicente}, J. and {Desai}, S. and {Diehl}, H.~T. and {Dietrich}, J.~P. and {Doel}, P. and {Drlica-Wagner}, A. and {Eckert}, K. and {Everett}, S. and {Ferrero}, I. and {Frieman}, J. and {Garc{\'\i}a-Bellido}, J. and {Gerdes}, D.~W. and {Giannantonio}, T. and {Gruendl}, R.~A. and {Gschwend}, J. and {Gutierrez}, G. and {Hartley}, W.~G. and {Hinton}, S.~R. and {Hollowood}, D.~L. and {Honscheid}, K. and {Hoyle}, B. and {Huff}, E.~M. and {Huterer}, D. and {Jain}, B. and {James}, D.~J. and {Jeltema}, T. and {Krause}, E. and {Kron}, R. and {Kuropatkin}, N. and {Lima}, M. and {Maia}, M.~A.~G. and {Marshall}, J.~L. and {Miquel}, R. and {Morgan}, R. and {Myles}, J. and {Palmese}, A. and {Paz-Chinch{\'o}n}, F. and {Rykoff}, E.~S. and {Samuroff}, S. and {Sanchez}, E. and {Scarpine}, V. and {Schubnell}, M. and {Serrano}, S. and {Sevilla-Noarbe}, I. and {Smith}, M. and {Soares-Santos}, M. and {Suchyta}, E. and {Swanson}, M.~E.~C. and {Tarle}, G. and {Thomas}, D. and {To}, C. and {Tucker}, D.~L. and {Varga}, T.~N. and {Wechsler}, R.~H. and {Weller}, J. and {Wester}, W. and {Wilkinson}, R.~D.},
        title = "{Dark energy survey year 3 results: weak lensing shape catalogue}",
      journal = {\mnras},
         year = 2021,
        month = jul,
       volume = {504},
       number = {3},
        pages = {4312-4336},
          doi = {10.1093/mnras/stab918},
archivePrefix = {arXiv},
       eprint = {2011.03408},
 primaryClass = {astro-ph.CO},
       adsurl = {https://ui.adsabs.harvard.edu/abs/2021MNRAS.504.4312G}
}

@ARTICLE{BPZ_2000,
       author = {{Ben{\'\i}tez}, Narciso},
        title = "{Bayesian Photometric Redshift Estimation}",
      journal = {\apj},
         year = 2000,
        month = jun,
       volume = {536},
       number = {2},
        pages = {571-583},
          doi = {10.1086/308947},
archivePrefix = {arXiv},
       eprint = {astro-ph/9811189},
 primaryClass = {astro-ph},
       adsurl = {https://ui.adsabs.harvard.edu/abs/2000ApJ...536..571B}
}

@ARTICLE{Sadeh_2016,
       author = {{Sadeh}, I. and {Abdalla}, F.~B. and {Lahav}, O.},
        title = "{ANNz2: Photometric Redshift and Probability Distribution Function Estimation using Machine Learning}",
      journal = {\pasp},
         year = 2016,
        month = oct,
       volume = {128},
       number = {968},
        pages = {104502},
          doi = {10.1088/1538-3873/128/968/104502},
archivePrefix = {arXiv},
       eprint = {1507.00490},
 primaryClass = {astro-ph.CO},
       adsurl = {https://ui.adsabs.harvard.edu/abs/2016PASP..128j4502S}
}

@ARTICLE{De_Vicente2016,
       author = {{De Vicente}, J. and {S{\'a}nchez}, E. and {Sevilla-Noarbe}, I.},
        title = "{DNF - Galaxy photometric redshift by Directional Neighbourhood Fitting}",
      journal = {\mnras},
         year = 2016,
        month = jul,
       volume = {459},
       number = {3},
        pages = {3078-3088},
          doi = {10.1093/mnras/stw857},
archivePrefix = {arXiv},
       eprint = {1511.07623},
 primaryClass = {astro-ph.CO},
       adsurl = {https://ui.adsabs.harvard.edu/abs/2016MNRAS.459.3078D}
}

@ARTICLE{DESY3_photometric,
       author = {{Sevilla-Noarbe}, I. and {Bechtol}, K. and {Carrasco Kind}, M. and {Carnero Rosell}, A. and {Becker}, M.~R. and {Drlica-Wagner}, A. and {Gruendl}, R.~A. and {Rykoff}, E.~S. and {Sheldon}, E. and {Yanny}, B. and {Alarcon}, A. and {Allam}, S. and {Amon}, A. and {Benoit-L{\'e}vy}, A. and {Bernstein}, G.~M. and {Bertin}, E. and {Burke}, D.~L. and {Carretero}, J. and {Choi}, A. and {Diehl}, H.~T. and {Everett}, S. and {Flaugher}, B. and {Gaztanaga}, E. and {Gschwend}, J. and {Harrison}, I. and {Hartley}, W.~G. and {Hoyle}, B. and {Jarvis}, M. and {Johnson}, M.~D. and {Kessler}, R. and {Kron}, R. and {Kuropatkin}, N. and {Leistedt}, B. and {Li}, T.~S. and {Menanteau}, F. and {Morganson}, E. and {Ogando}, R.~L.~C. and {Palmese}, A. and {Paz-Chinch{\'o}n}, F. and {Pieres}, A. and {Pond}, C. and {Rodriguez-Monroy}, M. and {Smith}, J. Allyn and {Stringer}, K.~M. and {Troxel}, M.~A. and {Tucker}, D.~L. and {de Vicente}, J. and {Wester}, W. and {Zhang}, Y. and {Abbott}, T.~M.~C. and {Aguena}, M. and {Annis}, J. and {Avila}, S. and {Bhargava}, S. and {Bridle}, S.~L. and {Brooks}, D. and {Brout}, D. and {Castander}, F.~J. and {Cawthon}, R. and {Chang}, C. and {Conselice}, C. and {Costanzi}, M. and {Crocce}, M. and {da Costa}, L.~N. and {Pereira}, M.~E.~S. and {Davis}, T.~M. and {Desai}, S. and {Dietrich}, J.~P. and {Doel}, P. and {Eckert}, K. and {Evrard}, A.~E. and {Ferrero}, I. and {Fosalba}, P. and {Garc{\'\i}a-Bellido}, J. and {Gerdes}, D.~W. and {Giannantonio}, T. and {Gruen}, D. and {Gutierrez}, G. and {Hinton}, S.~R. and {Hollowood}, D.~L. and {Honscheid}, K. and {Huff}, E.~M. and {Huterer}, D. and {James}, D.~J. and {Jeltema}, T. and {Kuehn}, K. and {Lahav}, O. and {Lidman}, C. and {Lima}, M. and {Lin}, H. and {Maia}, M.~A.~G. and {Marshall}, J.~L. and {Martini}, P. and {Melchior}, P. and {Miquel}, R. and {Mohr}, J.~J. and {Morgan}, R. and {Neilsen}, E. and {Plazas}, A.~A. and {Romer}, A.~K. and {Roodman}, A. and {Sanchez}, E. and {Scarpine}, V. and {Schubnell}, M. and {Serrano}, S. and {Smith}, M. and {Suchyta}, E. and {Tarle}, G. and {Thomas}, D. and {To}, C. and {Varga}, T.~N. and {Wechsler}, R.~H. and {Weller}, J. and {Wilkinson}, R.~D. and {DES Collaboration}},
        title = "{Dark Energy Survey Year 3 Results: Photometric Data Set for Cosmology}",
      journal = {\apjs},
         year = 2021,
        month = jun,
       volume = {254},
       number = {2},
          eid = {24},
        pages = {24},
          doi = {10.3847/1538-4365/abeb66},
archivePrefix = {arXiv},
       eprint = {2011.03407},
 primaryClass = {astro-ph.CO},
       adsurl = {https://ui.adsabs.harvard.edu/abs/2021ApJS..254...24S}
}

@article{Kilbinger_2014,
    author = "Kilbinger, Martin",
    title = "{Cosmology with cosmic shear observations: a review}",
    eprint = "1411.0115",
    archivePrefix = "arXiv",
    primaryClass = "astro-ph.CO",
    doi = "10.1088/0034-4885/78/8/086901",
    journal = "Rept. Prog. Phys.",
    volume = "78",
    pages = "086901",
    year = "2015"
}

@ARTICLE{Hinshaw_2013,
       author = {{Hinshaw}, G. and {Larson}, D. and {Komatsu}, E. and {Spergel}, D.~N. and {Bennett}, C.~L. and {Dunkley}, J. and {Nolta}, M.~R. and {Halpern}, M. and {Hill}, R.~S. and {Odegard}, N. and {Page}, L. and {Smith}, K.~M. and {Weiland}, J.~L. and {Gold}, B. and {Jarosik}, N. and {Kogut}, A. and {Limon}, M. and {Meyer}, S.~S. and {Tucker}, G.~S. and {Wollack}, E. and {Wright}, E.~L.},
        title = "{Nine-year Wilkinson Microwave Anisotropy Probe (WMAP) Observations: Cosmological Parameter Results}",
      journal = {\apjs},
         year = 2013,
        month = oct,
       volume = {208},
       number = {2},
          eid = {19},
        pages = {19},
          doi = {10.1088/0067-0049/208/2/19},
archivePrefix = {arXiv},
       eprint = {1212.5226},
 primaryClass = {astro-ph.CO},
       adsurl = {https://ui.adsabs.harvard.edu/abs/2013ApJS..208...19H}
}

@ARTICLE{Neyrinck_2008,
       author = {{Neyrinck}, Mark C.},
        title = "{ZOBOV: a parameter-free void-finding algorithm}",
      journal = {\mnras},
         year = 2008,
        month = jun,
       volume = {386},
       number = {4},
        pages = {2101-2109},
          doi = {10.1111/j.1365-2966.2008.13180.x},
archivePrefix = {arXiv},
       eprint = {0712.3049},
 primaryClass = {astro-ph},
       adsurl = {https://ui.adsabs.harvard.edu/abs/2008MNRAS.386.2101N}
}

@ARTICLE{sutter_2015,
       author = {{Sutter}, P.~M. and {Lavaux}, G. and {Hamaus}, N. and {Pisani}, A. and {Wandelt}, B.~D. and {Warren}, M. and {Villaescusa-Navarro}, F. and {Zivick}, P. and {Mao}, Q. and {Thompson}, B.~B.},
        title = "{VIDE: The Void IDentification and Examination toolkit}",
      journal = {Astronomy and Computing},
         year = 2015,
        month = mar,
       volume = {9},
        pages = {1-9},
          doi = {10.1016/j.ascom.2014.10.002},
archivePrefix = {arXiv},
       eprint = {1406.1191},
 primaryClass = {astro-ph.CO},
       adsurl = {https://ui.adsabs.harvard.edu/abs/2015A&C.....9....1S}
}

@ARTICLE{Takahashi_2017,
       author = {{Takahashi}, Ryuichi and {Hamana}, Takashi and {Shirasaki}, Masato and {Namikawa}, Toshiya and {Nishimichi}, Takahiro and {Osato}, Ken and {Shiroyama}, Kosei},
        title = "{Full-sky Gravitational Lensing Simulation for Large-area Galaxy Surveys and Cosmic Microwave Background Experiments}",
      journal = {\apj},
         year = 2017,
        month = nov,
       volume = {850},
       number = {1},
          eid = {24},
        pages = {24},
          doi = {10.3847/1538-4357/aa943d},
archivePrefix = {arXiv},
       eprint = {1706.01472},
 primaryClass = {astro-ph.CO},
       adsurl = {https://ui.adsabs.harvard.edu/abs/2017ApJ...850...24T}
}

@article{Cautun2014,
    author = {Cautun, Marius and van de Weygaert, Rien and Jones, Bernard J. T. and Frenk, Carlos S.},
    title = "{Evolution of the cosmic web}",
    journal = {Mon. Not. Roy. Astron. Soc.},
    volume = {441},
    number = {4},
    pages = {2923-2973},
    year = {2014},
    month = {05},
    issn = {0035-8711},
    doi = {10.1093/mnras/stu768},
    url = {https://doi.org/10.1093/mnras/stu768},
    eprint = {http://oup.prod.sis.lan/mnras/article-pdf/441/4/2923/4046922/stu768.pdf},
}

@ARTICLE{Jeffrey2021,
       author = {{Jeffrey}, N. and {Gatti}, M. and {Chang}, C. and {Whiteway}, L. and {Demirbozan}, U. and {Kovacs}, A. and {Pollina}, G. and {Bacon}, D. and {Hamaus}, N. and {Kacprzak}, T. and {Lahav}, O. and {Lanusse}, F. and {Mawdsley}, B. and {Nadathur}, S. and {Starck}, J.~L. and {Vielzeuf}, P. and {Zeurcher}, D. and {Alarcon}, A. and {Amon}, A. and {Bechtol}, K. and {Bernstein}, G.~M. and {Campos}, A. and {Rosell}, A. Carnero and {Kind}, M. Carrasco and {Cawthon}, R. and {Chen}, R. and {Choi}, A. and {Cordero}, J. and {Davis}, C. and {DeRose}, J. and {Doux}, C. and {Drlica-Wagner}, A. and {Eckert}, K. and {Elsner}, F. and {Elvin-Poole}, J. and {Everett}, S. and {Fert{\'e}}, A. and {Giannini}, G. and {Gruen}, D. and {Gruendl}, R.~A. and {Harrison}, I. and {Hartley}, W.~G. and {Herner}, K. and {Huff}, E.~M. and {Huterer}, D. and {Kuropatkin}, N. and {Jarvis}, M. and {Leget}, P.~F. and {MacCrann}, N. and {McCullough}, J. and {Muir}, J. and {Myles}, J. and {Navarro-Alsina}, A. and {Pandey}, S. and {Prat}, J. and {Raveri}, M. and {Rollins}, R.~P. and {Ross}, A.~J. and {Rykoff}, E.~S. and {S{\'a}nchez}, C. and {Secco}, L.~F. and {Sevilla-Noarbe}, I. and {Sheldon}, E. and {Shin}, T. and {Troxel}, M.~A. and {Tutusaus}, I. and {Varga}, T.~N. and {Yanny}, B. and {Yin}, B. and {Zhang}, Y. and {Zuntz}, J. and {Abbott}, T.~M.~C. and {Aguena}, M. and {Allam}, S. and {Andrade-Oliveira}, F. and {Becker}, M.~R. and {Bertin}, E. and {Bhargava}, S. and {Brooks}, D. and {Burke}, D.~L. and {Carretero}, J. and {Castander}, F.~J. and {Conselice}, C. and {Costanzi}, M. and {Crocce}, M. and {da Costa}, L.~N. and {Pereira}, M.~E.~S. and {De Vicente}, J. and {Desai}, S. and {Diehl}, H.~T. and {Dietrich}, J.~P. and {Doel}, P. and {Ferrero}, I. and {Flaugher}, B. and {Fosalba}, P. and {Garc{\'\i}a-Bellido}, J. and {Gaztanaga}, E. and {Gerdes}, D.~W. and {Giannantonio}, T. and {Gschwend}, J. and {Gutierrez}, G. and {Hinton}, S.~R. and {Hollowood}, D.~L. and {Hoyle}, B. and {Jain}, B. and {James}, D.~J. and {Lima}, M. and {Maia}, M.~A.~G. and {March}, M. and {Marshall}, J.~L. and {Melchior}, P. and {Menanteau}, F. and {Miquel}, R. and {Mohr}, J.~J. and {Morgan}, R. and {Ogando}, R.~L.~C. and {Palmese}, A. and {Paz-Chinch{\'o}n}, F. and {Plazas}, A.~A. and {Rodriguez-Monroy}, M. and {Roodman}, A. and {Sanchez}, E. and {Scarpine}, V. and {Serrano}, S. and {Smith}, M. and {Soares-Santos}, M. and {Suchyta}, E. and {Tarle}, G. and {Thomas}, D. and {To}, C. and {Weller}, J. and {DES Collaboration}},
        title = "{Dark Energy Survey Year 3 results: Curved-sky weak lensing mass map reconstruction}",
      journal = {\mnras},
         year = 2021,
        month = aug,
       volume = {505},
       number = {3},
        pages = {4626-4645},
          doi = {10.1093/mnras/stab1495},
archivePrefix = {arXiv},
       eprint = {2105.13539},
 primaryClass = {astro-ph.CO},
       adsurl = {https://ui.adsabs.harvard.edu/abs/2021MNRAS.505.4626J}
}

@ARTICLE{Chang2018,
       author = {{Chang}, C. and {Pujol}, A. and {Mawdsley}, B. and {Bacon}, D. and {Elvin-Poole}, J. and {Melchior}, P. and {Kov{\'a}cs}, A. and {Jain}, B. and {Leistedt}, B. and {Giannantonio}, T. and {Alarcon}, A. and {Baxter}, E. and {Bechtol}, K. and {Becker}, M.~R. and {Benoit-L{\'e}vy}, A. and {Bernstein}, G.~M. and {Bonnett}, C. and {Busha}, M.~T. and {Carnero Rosell}, A. and {Castander}, F.~J. and {Cawthon}, R. and {da Costa}, L.~N. and {Davis}, C. and {De Vicente}, J. and {DeRose}, J. and {Drlica-Wagner}, A. and {Fosalba}, P. and {Gatti}, M. and {Gaztanaga}, E. and {Gruen}, D. and {Gschwend}, J. and {Hartley}, W.~G. and {Hoyle}, B. and {Huff}, E.~M. and {Jarvis}, M. and {Jeffrey}, N. and {Kacprzak}, T. and {Lin}, H. and {MacCrann}, N. and {Maia}, M.~A.~G. and {Ogando}, R.~L.~C. and {Prat}, J. and {Rau}, M.~M. and {Rollins}, R.~P. and {Roodman}, A. and {Rozo}, E. and {Rykoff}, E.~S. and {Samuroff}, S. and {S{\'a}nchez}, C. and {Sevilla-Noarbe}, I. and {Sheldon}, E. and {Troxel}, M.~A. and {Varga}, T.~N. and {Vielzeuf}, P. and {Vikram}, V. and {Wechsler}, R.~H. and {Zuntz}, J. and {Abbott}, T.~M.~C. and {Abdalla}, F.~B. and {Allam}, S. and {Annis}, J. and {Bertin}, E. and {Brooks}, D. and {Buckley-Geer}, E. and {Burke}, D.~L. and {Carrasco Kind}, M. and {Carretero}, J. and {Crocce}, M. and {Cunha}, C.~E. and {D'Andrea}, C.~B. and {Desai}, S. and {Diehl}, H.~T. and {Dietrich}, J.~P. and {Doel}, P. and {Estrada}, J. and {Fausti Neto}, A. and {Fernandez}, E. and {Flaugher}, B. and {Frieman}, J. and {Garc{\'\i}a-Bellido}, J. and {Gruendl}, R.~A. and {Gutierrez}, G. and {Honscheid}, K. and {James}, D.~J. and {Jeltema}, T. and {Johnson}, M.~W.~G. and {Johnson}, M.~D. and {Kent}, S. and {Kirk}, D. and {Krause}, E. and {Kuehn}, K. and {Kuhlmann}, S. and {Lahav}, O. and {Li}, T.~S. and {Lima}, M. and {March}, M. and {Martini}, P. and {Menanteau}, F. and {Miquel}, R. and {Mohr}, J.~J. and {Neilsen}, E. and {Nichol}, R.~C. and {Petravick}, D. and {Plazas}, A.~A. and {Romer}, A.~K. and {Sako}, M. and {Sanchez}, E. and {Scarpine}, V. and {Schubnell}, M. and {Smith}, M. and {Smith}, R.~C. and {Soares-Santos}, M. and {Sobreira}, F. and {Suchyta}, E. and {Tarle}, G. and {Thomas}, D. and {Tucker}, D.~L. and {Walker}, A.~R. and {Wester}, W. and {Zhang}, Y. and {DES Collaboration}},
        title = "{Dark Energy Survey Year 1 results: curved-sky weak lensing mass map}",
      journal = {\mnras},
         year = 2018,
        month = apr,
       volume = {475},
       number = {3},
        pages = {3165-3190},
          doi = {10.1093/mnras/stx3363},
archivePrefix = {arXiv},
       eprint = {1708.01535},
 primaryClass = {astro-ph.CO},
       adsurl = {https://ui.adsabs.harvard.edu/abs/2018MNRAS.475.3165C}
}

\end{document}